\documentclass[aps,prb,twocolumn]{revtex4}
\usepackage{color}
\usepackage{float}
\usepackage{amsmath}
\usepackage{amssymb}
\usepackage{graphicx}
\newcommand{\reff}[1]{(\ref{#1})}
\begin{document}
\title{Charging graphene nanoribbon quantum dots}
\author{D. P. \.Zebrowski and B. Szafran}
\affiliation{AGH University of Science and Technology, Faculty of Physics and Applied Computer Science,\\
al. Mickiewicza 30, 30-059 Krak\'ow, Poland}
\begin{abstract}
We describe charging a quantum dot induced electrostatically within a semiconducting graphene nanoribbon by electrons or holes.
The applied model is based on a tight-binding approach with the electron-electron interaction introduced by a mean field local spin density approximation.
The numerical approach accounts for the charge of all the $p_z$ electrons and screening of external potentials by states near the charge neutrality point.
Both a homogenous ribbon and a graphene flake embedded within the ribbon are discussed.
The formation of transport gaps as functions of the external confinement potential (top gate potential) and the Fermi energy (back gate potential)  are described in good qualitative agreement with the experimental data.
For a fixed number of excess electrons we find that the excess charge added to the system is, - depending on the voltages defining the work point of the device:
{\it i)} delocalized outside the quantum dot -- in the transport gap due to the top gate potential {\it ii)} localized inside the quantum dot -- in the transport gap due to the back gate potential or {\it iii)}
extended over both the
quantum dot area and the ribbon connections -- outside the transport gaps.  The applicability of the frozen valence band approximation to describe charging the quantum dot by excess electrons is also discussed.
\end{abstract}
\maketitle

\section{Introduction}

The  dispersion relation of graphene \cite{Neto2009} -- gapless and linear near the charge neutrality point --
prevents electrostatic confinement of charge carriers \cite{Klein29,Young}.
The energy gap in the band structure is opened for finite strips of graphene \cite{nrb,waka,nrbr,nrb4}
which allows for carrier confinement by external potentials defining quantum dots inside the  nanoribbons \cite{silvestrov07,han,nrb3,trau,oostinga,xliu,stampfer,moreview,droescher}.
All nanoribbons exhibit a transport gap \cite{han,nrb2} near the neutrality point.
Formation of the transport gap can be described theoretically already for non-interacting electrons as due to lateral confinement and edge disorder \cite{evaldsson}.
Nevertheless, the Anderson localization effects are reinforced by the electron-electron interaction
and according to the present understanding, the transport gap is stabilized by a spontaneous formation of multiple Coulomb islands along the ribbon.
The current peaks that appear for a biased ribbon inside the transport gap are attributed to electrons hopping between the charge puddles \cite{sols,stampfer,xliu,nrb3,chiu,mhan,todd,gala}.
Charging of the intentionally defined quantum dots appears also within the transport gap \cite{chiu,stampfer,moreview,droescher}.
The spontaneous disorder-induced quantum dots  and the ones defined electrostatically are characterized by Coulomb diamonds in the charge stability diagrams \cite{chiu,stampfer,xliu}.
In part of the studies the nanoribbons contain an additional inline graphene flake \cite{stampfer,moreview, gut,chiu,sch,gutti} in the gated region, which is useful
in separation\cite{chiu} of the charging effects due to the intentionally introduced quantum dot and the spontaneous Coulomb islands along the ribbon.

The electronic properties of quantum dots made of finite graphene flakes have been extensively studied in a number of papers \cite{Ezawa,Potasz1,Yamamoto,Zhang,Guclu,Potasz2,Palacios,Wang,Zarenia,Wunsch}
and only recently a finite graphene flake was successfully connected directly to source-drain electrodes without nanoribbons feeding the charge to the flake \cite{mori}.
Localization of electrons within a quantum dot induced electrostatically inside a graphene sheet with the mass gap induced by the substrate
have been studied in Ref. \onlinecite{rontani}, and the one-electron properties of these dots were discussed in Refs. \onlinecite{these1,these2}.

The purpose of the present paper is a simulation of charging a quantum dot defined by an external potential within a graphene nanoribbon.
We consider both  homogenous ribbons and the ones containing an inline flake of graphene. We prepared a numerical model that covers formation of both n-type and p-type quantum dots confining
conduction band electrons or valence bands holes depending on the sign of the external potential. For the purpose of the present
study we consider a semiconducting armchair nanoribbon using the tight-binding approach for the single-electron spectrum and the LSDA potentials
describing the electron-electron interaction. The model accounts for contribution of all the $p_z$ electrons within the sample in formation of the potential profile along the system.
We find that within the transport gap the quantum dot charging in terms of the chemical potentials can be very well described in a single-electron basis limited to an energy range near the neutrality point.
The charge density of the states deep inside the valence band can then be considered frozen without any detectable deviation from the electron-hole symmetry in terms
of electron and hole charging processes.
We discuss the transport gap and the charge distribution within the entire system including an integer quantization of the quantum-dot-confined charge.
For a neutral system the modification of the lateral confinement - with introduction of the flake into the ribbon --
does not induce formation of a localized charge puddle. Nevertheless, the excess electrons -- when present within the system -- get localized within the flake already in the absence of the external potential.
A good qualitative correspondence with the transport gaps observed in Ref. \cite{xliu} as functions of the back gate and the top gate potentials is found.
Applicability of the approximation of a frozen valence band for charging the dot with excess electrons is also discussed.
%We find that the system confined within the quantum dot forms barriers of the effective potential which separates the confined charge puddle from the rest of the system.

\begin{figure}[ht]
\begin{center}
\begin{tabular}{ll}
a) \includegraphics[angle=0,width=0.17\textwidth]{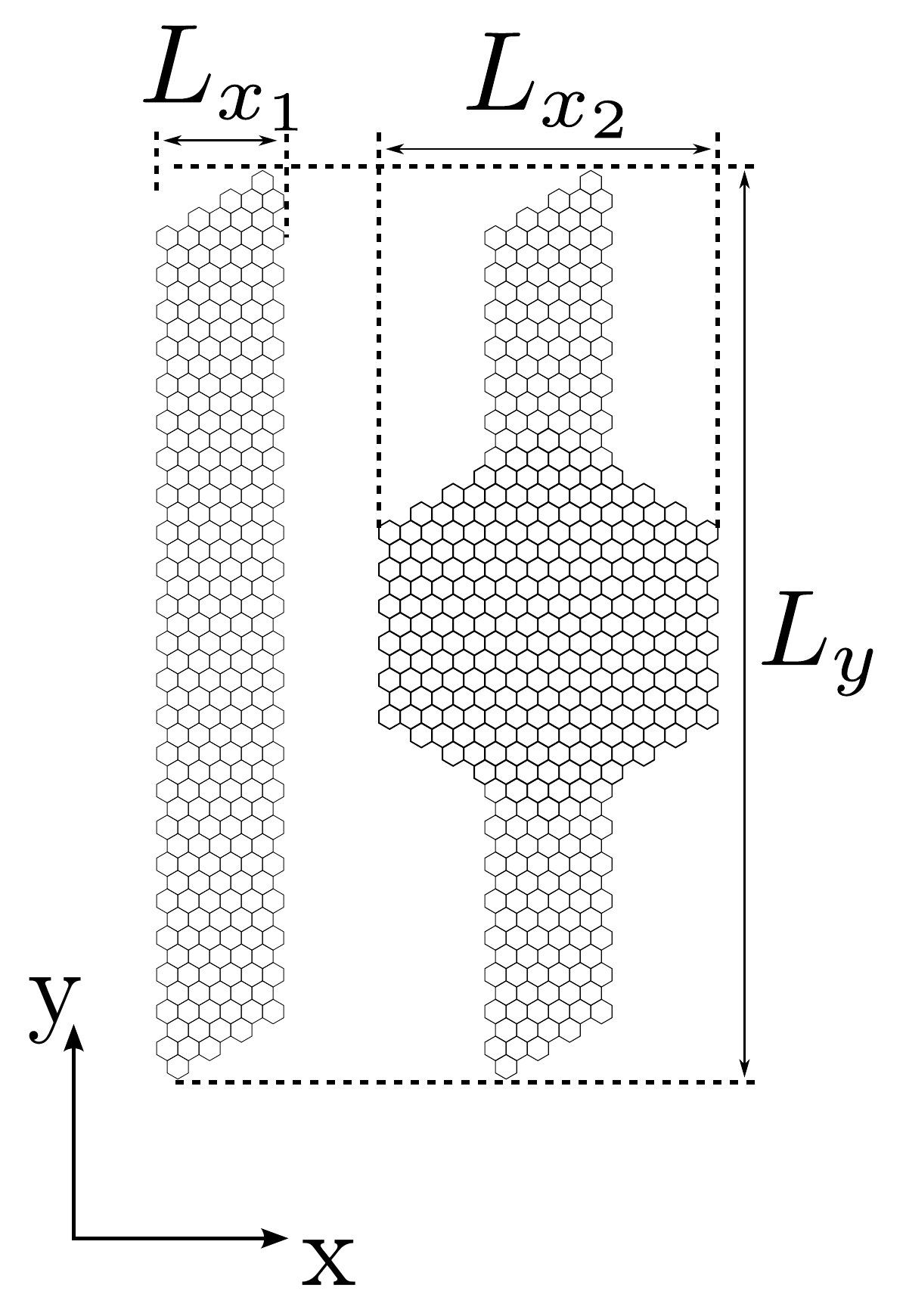}& b)\hspace{-0.6cm} \includegraphics[angle=0,width=0.28\textwidth]{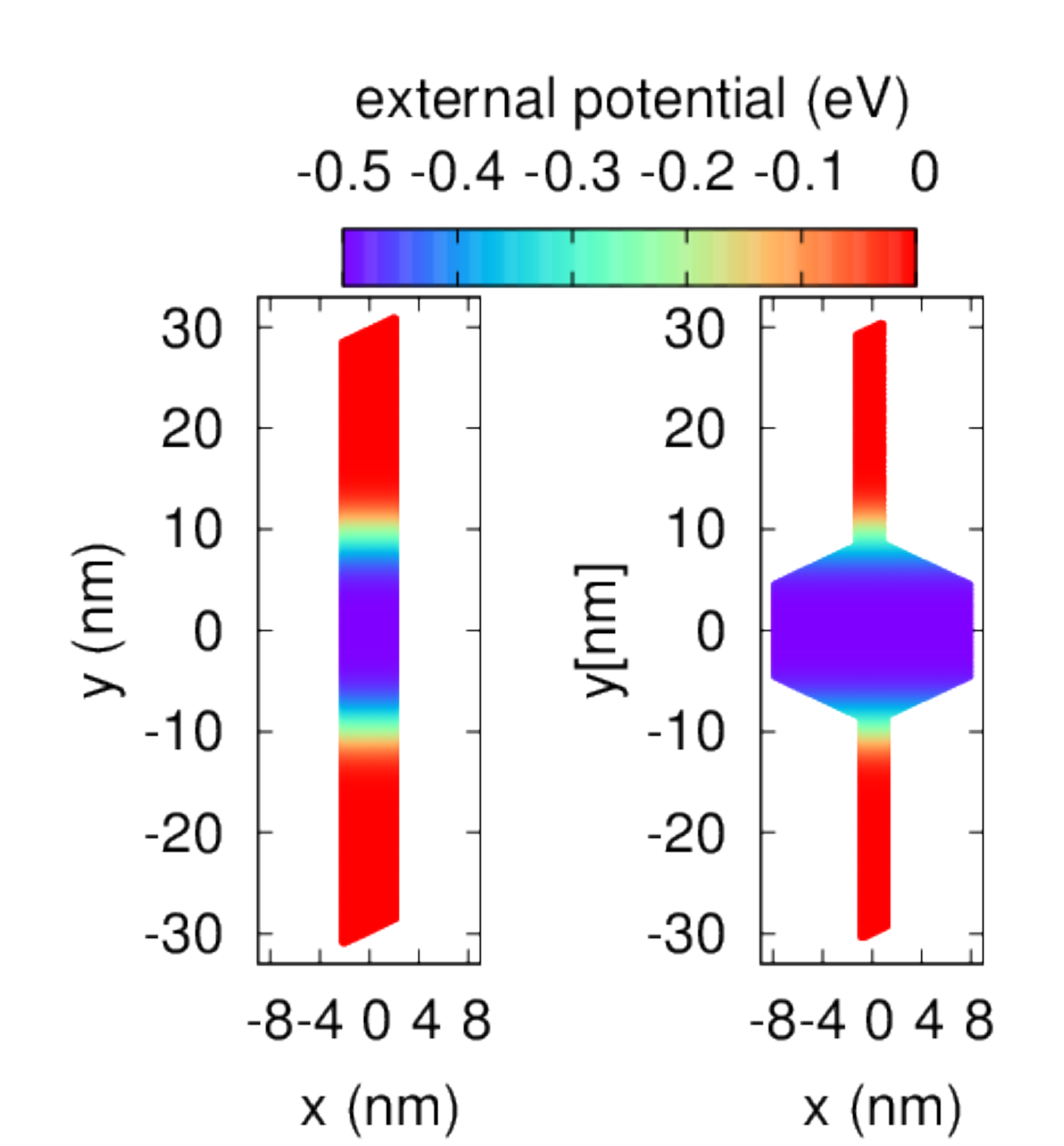}
 \end{tabular}
\caption{(a) Schematics of the ribbons considered in this work with and without the central flake.
The considered nanoribbons have armchair edges and the number of atoms across the channel (18 and 36) corresponds to a semiconducting
dispersion relation of $p_z$ tight-binding Hamiltonian. The length of the ribbon is $L_y=61$ nm.
The hexagonal flake of width $L_{x_2}=15.7$ nm is embedded inside the ribbon the width $L_{x_1}=2.15$ nm (left plot)
We consider also homogenous ribbons of width $L_{x_1}=4.3$ nm (left plot).
(b) External potential defining the quantum dot as given by Eq. \reff{Vext} for $W=0.5$ eV.
}\label{u}
\end{center}
\end{figure}

\section{Theory}
We consider a finite ribbon with armchair edges [see Fig. \ref{u}(a)] with or without a hexagonal flake defined in its center.
We use the single-electron Hamiltonian as given by the tight-binding method for $p_z$ orbitals
\begin{equation}
 \hat{H}=\sum_{<i,i'>} t \hat{c}^{\dagger}_i \hat{c}_{i'}+h.c.,
 \label{H1}
\end{equation}
where $t=-2.7$ eV is the hopping parameter, and $\hat{c}^{\dagger}_i$ is the operator of electron creation at $i$-th atom and the summation runs over nearest neighbor atoms.

Upon diagonalization of the Hamiltonian  \label{H1} we obtain the eigenstates of form
\begin{equation}
 \psi_j(\mathbf{r})=\sum_{i=1}^{N_a}c_{ij} p_z(\mathbf{r-r}_i).
 \label{ffal}
\end{equation}
We assume 18 or 36 atoms across the  ribbon, so that the system -- in the $p_z$ tight-binding approach
has a semiconducting dispersion relation [see Fig. 1].
The charge redistribution that we describe requires application of large external potentials, for which the potential at the ions is far from the charge neutrality point and the linear dispersion relation of graphene. For this reason we need to employ an atomistic tight-binding approach instead of its continuous low-energy approximation in the Dirac form. The atomistic approach has its price on computational complexity and memory consumption, that limits the size of the systems than can be studied.   Nevertheless,  the obtained picture is qualitatively independent of the size the systems and the ones considered in this work remain within an experimental reach. For applications of the graphene nanoribbons in electronics their width below 10 nm is needed to switch off the current  \cite{cyt1,cyt2}.  The external potential well defining the size of the quantum dot is 15-20 nm [Fig. 1(b)], while the estimated diameter of the quantum dot  in a gated system is 25-40 nm \cite{mori}.

The system of interacting electrons is treated by a mean field approach.
The number of ions in the considered systems and the number of $p_z$ orbitals in the basis ranges up to about $N_a=15$ thousands, which sets the size of the basis \reff{ffal}.
The considered systems are close to the charge neutrality,
with the number of electrons within the ribbon close to the number of positive ions  (net charge $\pm 24e$).
We have found that the ground-state properties of the system can be quite accurately described using a basis of a few hundred wave functions
of type (\ref{ffal}) for the energies near the neutrality point.

We order the single-electron basis functions by the energy eigenvalues $\varepsilon _j$, where $j$ is an integer non-zero index
$j\in[-N_a/2,N_a/2]$ ($N_a$ is an even number).
For the LSDA \cite{Perdew} calculations we use a basis of form
\begin{equation}
 \phi^{\sigma}_k(\mathbf{r})=\sum_{j=-m}^{m} d^{\sigma}_{jk} \psi_j(\mathbf{r}),
 \label{nbaza}
\end{equation}
where $\sigma=\uparrow \downarrow $ is the spin index. The basis \reff{nbaza} contains $m$ single-electron states of both the valence (negative $j$) and the conduction bands (positive $j$ index).
The states that are deep below the energy gap with $j\in [-N_a/2,-m-1]$ are considered
frozen. Their contribution to the electron spin density is calculated once and for all after diagonalization of the single-electron Hamiltonian
\begin{equation}
 n^{\sigma}_{fix}(\mathbf{r})=\sum_{j=-N_a/2}^{-m-1} |\psi_j(\mathbf{r})|^2. \label{frozen}
\end{equation}
For the simulation of the mean-field potential we neglect the overlap of neighbor orbitals;
the approach is equivalent to two-center treatment of the Coulomb integrals \cite{2center}.
Accordingly, the contribution of the $j-$ th orbital to the spin density at $l$-th ion is given by $|c_{lj}p_z({\bf r}-{\bf r}_j)|^2$, and
\begin{equation}
  n^{\sigma}_{fix}(\mathbf{r_l})=\sum_{j=-N_a/2}^{-m-1} |\psi_j(\mathbf{r_l})|^2=\sum_{j=-N_a/2}^{-m-1} |c_{lj}p_z({\bf r}-{\bf r}_j)|^2.
\end{equation}

The electron-electron interaction is introduced by the LSDA mean field potential \cite{Perdew},
with the Hamiltonian of form
\begin{equation}
 \hat{H}^{\sigma}_{DFT}=\sum_{j}\hat{d}^{\sigma\dagger}_j \hat{d}^{\sigma}_{j}\epsilon_j^\sigma+\sum_{j,j'} \tau^{\sigma}_{jj'} \hat{d}^{\sigma\dagger}_j \hat{d}^{\sigma}_{j'},\label{dft}
\end{equation}
where $\hat{d}^{\sigma\dagger}_{j}$ is the electron creation operator in the  eigenstate $j$ of Hamiltonian (1) with spin $\sigma$.
The DFT potential enters the hopping parameters ($\tau^{\sigma}_{jj'} $),
\begin{equation}
 \tau^{\sigma}_{jj'}=\int\psi^*_{j}(\mathbf{r})V^{\sigma}(\mathbf{r},n^{\uparrow}(\mathbf{r}),n^{\downarrow}(\mathbf{r}))\psi_{j'}(\mathbf{r}) d^3\mathbf{r} \label{tautau}
\end{equation}
where $ n^{\sigma}(\mathbf{r})$ is the spin $\sigma$ density
\begin{equation}
 n^{\sigma}(\mathbf{r})=n'^\sigma({\mathbf{r}})+n^{\sigma}_{fix}(\mathbf{r}).
  \label{gestoscf}
\end{equation}
In the above formula $n'^\sigma$ is the spin density of electrons occupying energy levels close to the neutrality point
which are determined by DFT,
\begin{equation}
n'^{\sigma}(\mathbf{r})=\sum_{k=-m}^{m}f(E_F,E^{\sigma}_k)|\phi_k^{\sigma}(\mathbf{r})|^2,
\end{equation}
where $f(E_F,E^{\sigma}_k)$ is the Fermi-Dirac distribution for the Fermi energy $E_F$,
and $E^\sigma_k$ are the eigenvalues of the  DFT energy operator \reff{dft}.
The Fermi energy $E_F$ is found from the normalization for the electron density $n({\mathbf{r}})=n^\uparrow(\mathbf{r})+n^\downarrow(\mathbf{r})$,
\begin{equation}
 \int n(\mathbf{r}) d^3\mathbf{r}=N,
\end{equation}
where $N$ is the number of electrons within the system $N=N_i+N_e$,
with $N_i$ that stands for the number of ions within the nanoribbon, and $N_e$ for the number of excess electrons with respect to a charge neutral system.

The potential $V^{\sigma}$ defining the hopping parameters \reff{tautau} is given by
\begin{equation}
 V^{\sigma}(\mathbf{r},n^{\uparrow}(\mathbf{r}),n^{\downarrow}(\mathbf{r}))=V_{ext}(\mathbf{r})+V_{H}(\mathbf{r},n)+V_{C}(\mathbf{r})+V^{\sigma}_{xc}(n^{\uparrow},n^{\downarrow}).
 \label{potW}
\end{equation}
The external electrostatic potential in Eq. \reff{potW}  defining the quantum dot in the center of the ribbon
is modelled by
\begin{equation}
 V_{ext}(\mathbf{r})=-W\exp\left(-\left(\frac{y}{R} \right)^4\right),\label{Vext}
\end{equation}
where $2R=20$ nm is the length of the quantum dot -- see Fig. \ref{u}(b) for the profile plotted for $W=0.5$ eV.
The electrostatic potential defining the quantum dot is assumed with an axial symmetry, since the gates are usually applied perpendicular to the nanoribbon \cite{xliu}. The potential as induced electrostatically is smooth and - due to the screening- short range. The exact profile of the confinement potential depends on the size of the gates and their distance to the area occupied by electrons \cite{bednarek1,bednarek2}. The qualitative features for the charging of the quantum dots remain independent of the potential profile, as long as it is smooth and short range. The results for a modified form of the potential is given in Section IV.E, with power of 2 instead of 4 in Eq.(12).

The Hartree potential  $V_H(\mathbf{r},n(\mathbf{r}))=\int \frac{en(\mathbf{r'})}{4\pi\epsilon\epsilon_0|\mathbf{r}-\mathbf{r'}|}d^3\mathbf{r'}$,
is evaluated as
% \begin{equation}
%   V_H(\mathbf{r}_l)=\sum_{i\neq l}\left(\frac{en_i}{4\pi\epsilon\epsilon_0|\mathbf{r}_i-\mathbf{r}_l|}\right)+n_i\int \frac{ep_z^2(\mathbf{r}-\mathbf{r}_i)}{4\pi\epsilon\epsilon_0|\mathbf{r}-\mathbf{r}_i|}d^3\mathbf{r},
%   \label{hpott}
% \end{equation}
\begin{equation}
  V_H(\mathbf{r}_l)=\frac{e^2}{4\pi\epsilon\epsilon_0} \left( \sum_{i\neq l}\frac{n_i}{|\mathbf{r}_i-\mathbf{r}_l|}+n_l\int \frac{p_z^2(\mathbf{r}-\mathbf{r}_l)}{|\mathbf{r}-\mathbf{r}_l|}d^3\mathbf{r}\right),
  \label{hpott}
\end{equation}
where the summation runs over the ions $i$, and  $n^{}(\mathbf{r_l})=n^{}_l$.
We use the dielectric constant $\epsilon=6$ as for the graphene grown on SiC \cite{sicc}.

In Eq. \reff{potW} $V_C$ is the potential of the carbon ions which is calculated in a similar manner
\begin{equation}
  V_C(\mathbf{r}_l)=-\frac{e^2}{4\pi\epsilon\epsilon_0}\left( \sum_{i\neq l}\frac{1}{|\mathbf{r}_i-\mathbf{r}_l|}+\int \frac{p_z^2(\mathbf{r}-\mathbf{r}_l)}{|\mathbf{r}-\mathbf{r}_l|}d^3\mathbf{r}\right),
  \label{hpott}
\end{equation}

The exchange and correlation potentials are taken according to the Perdew-Zunger parametrization \cite{Perdew}
for the spin density given at the ions.
Once, the potential are known, we calculate the hopping elements (\ref{tautau})
 \begin{equation}
 \tau^{\sigma}_{jj'}=\sum_{i=1}^{N_a}V_{i}^{\sigma}c^*_{ij}c_{ij'}
\end{equation}
where $ V_{i}^{\sigma}=V^{\sigma}(\mathbf{r}_i,n^{\uparrow},n^{\downarrow})$

We evaluate the chemical potential of the $N$ electron system
\begin{equation}
 \mu_{N}=E^{tot}_N-E^{tot}_{N-1},
\end{equation}
where $E^{tot}_N$ is the total energy,
calculated as
\begin{equation}
 E^{tot}_N=T+E_{ext,C}+E_{xc}+E_{H},\label{etot}
\end{equation}
with the kinetic energy
\begin{equation}
 T=-\frac{\hbar^2}{2m_0}\sum_{\sigma=\uparrow,\downarrow}\sum_{k=-m}^{m}f(E_f,E^{\sigma}_k)\int \left ( \phi^{\sigma}_k(\mathbf{r})\right)^* \nabla^2 \phi^{\sigma}_k(\mathbf{r}) d^3\mathbf{r},\label{t}
\end{equation}
which can be expressed using the single-electron energy eigenvalues $\varepsilon_j$ for operator (1) with eigenfunctions \reff{nbaza}
\begin{equation}
 T=\sum_{\sigma=\uparrow\downarrow}\sum_{k=-m}^{m}f(E_f,E^{\sigma}_k)\sum_{j=-m}^{m}|d_{jk}|^2\varepsilon_j.
\end{equation}
$E_{ext,C}$ in Eq. \reff{etot} is the contribution to the energy of the external potential and the carbon lattice
\begin{equation}
 E_{ext,C}=\int (V_{ext}(\mathbf{r})+V_{C}(\mathbf{r}))n(\mathbf{r}) d^3\mathbf{r}.
\end{equation}
$E_{xc}$ in Eq. (17) is the Perdew-Zunger \cite{Perdew} exchange-correlation energy and the last term of \reff{etot} is
the Hartree energy
\begin{equation}
 E_{H}=\frac{1}{2}\int V_{H}(\mathbf{r})n(\mathbf{r}) d^3\mathbf{r}.
\end{equation}
In the summation of the single-electron contributions to the kinetic energy $T$ \reff{t} we account only for the $2m$ orbitals forming the
basis for the DFT calculation, since the contributions of lower energy levels -- considered frozen -- cancel
in the evaluation of the chemical potential anyway. However, in the exchange-correlation and Hartree potential, the
total electron density, including the frozen orbitals needs to be taken into account.
For convergence of the DFT equations we use the Broyden\cite{broyden} method.
The convergence is further additionally enhanced by annealing of the temperature. We start
from 20K and take down the temperature to $T= 10$ K. This range of the temperature corresponds to the
one actually used in the experimental studies of graphene quantum dot \cite{todd}.
In the model system, which is small, below 10 K no further potential dependence on the temperature is observed.

\begin{figure}[htbp]
\begin{center}
\begin{tabular}{ll}
a) & \includegraphics[angle=0,width=0.45\textwidth]{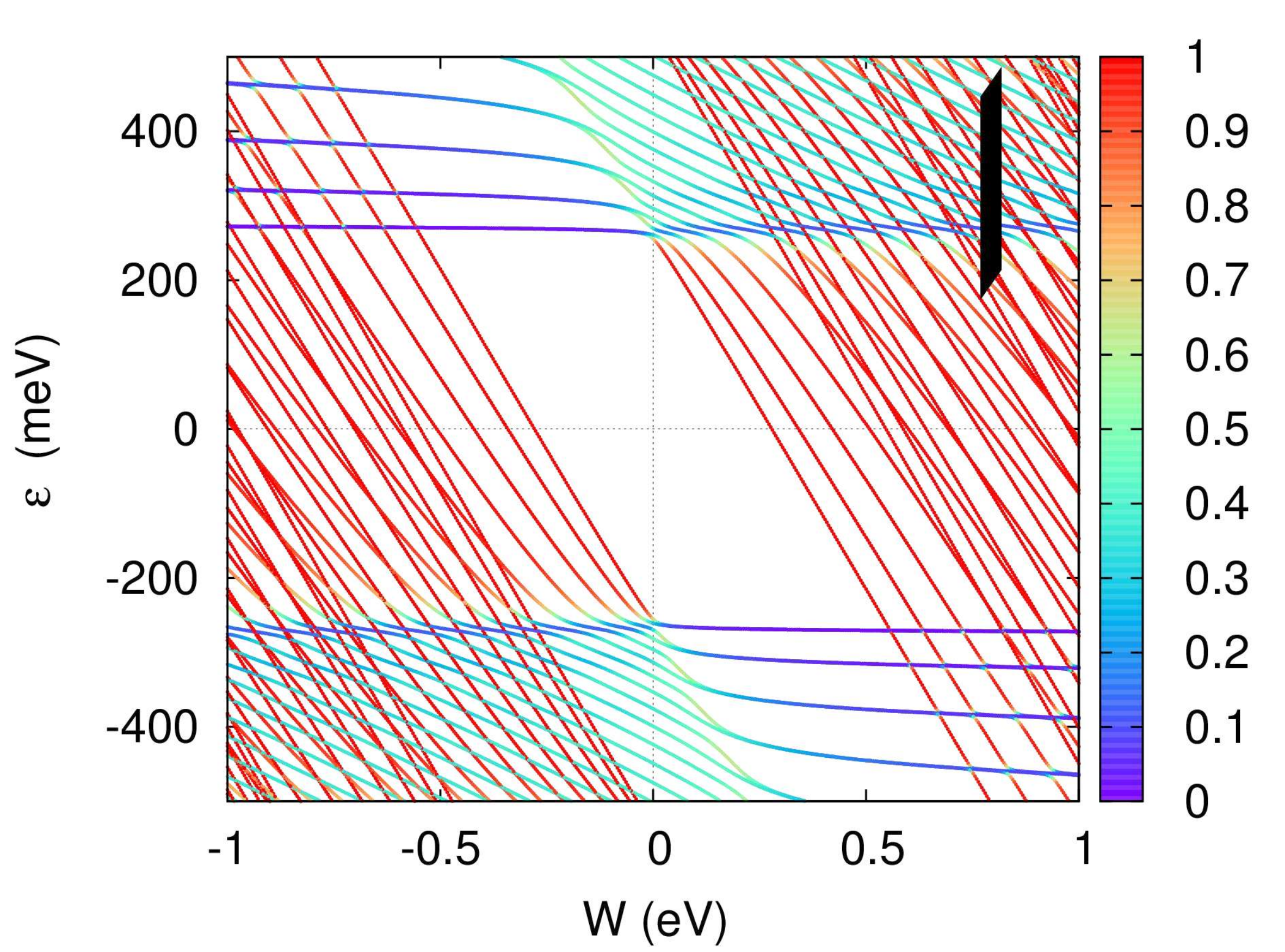} \\
b) & \includegraphics[angle=0,width=0.45\textwidth]{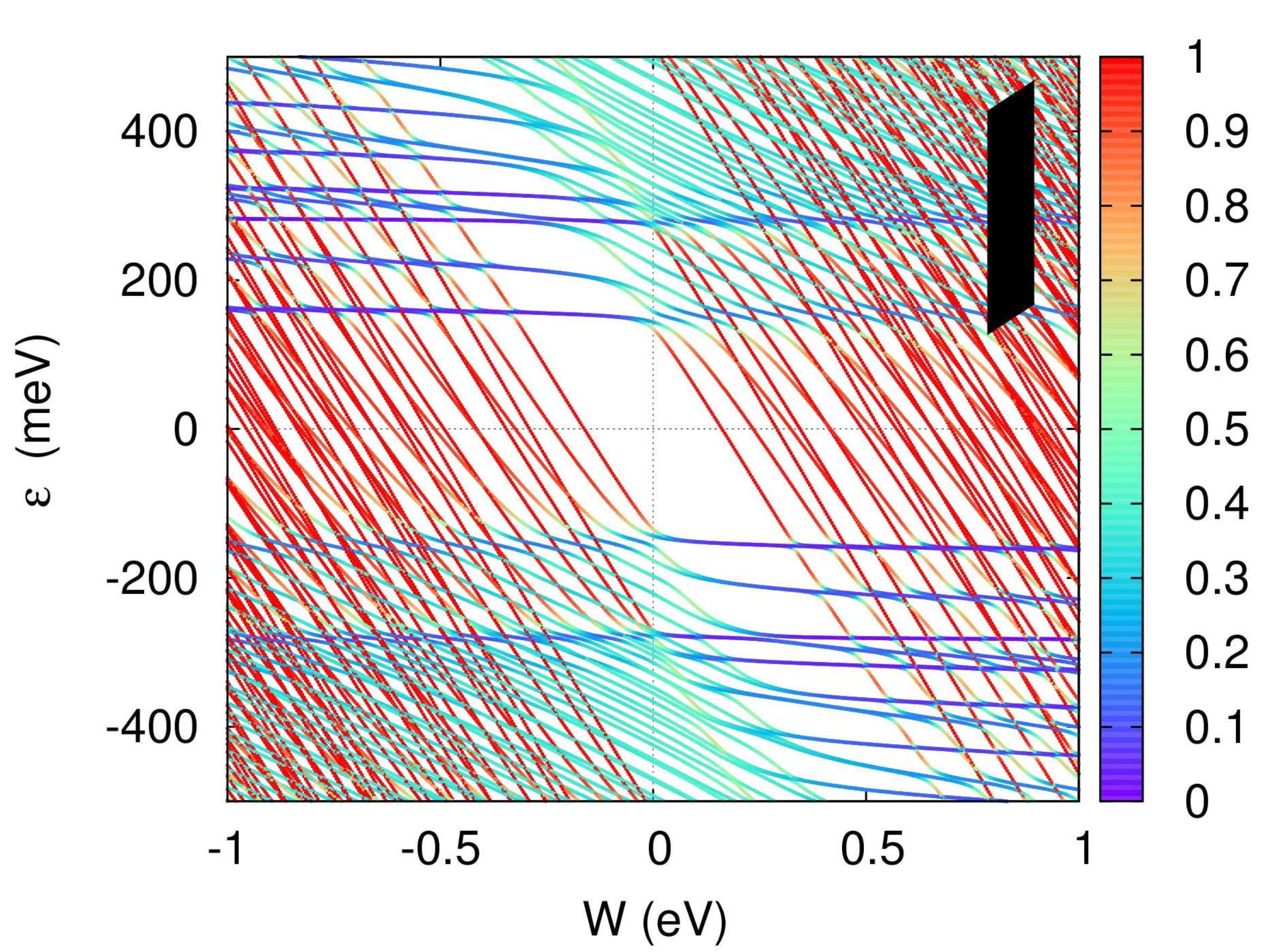} \\
c)&\includegraphics[angle=0,width=0.45\textwidth]{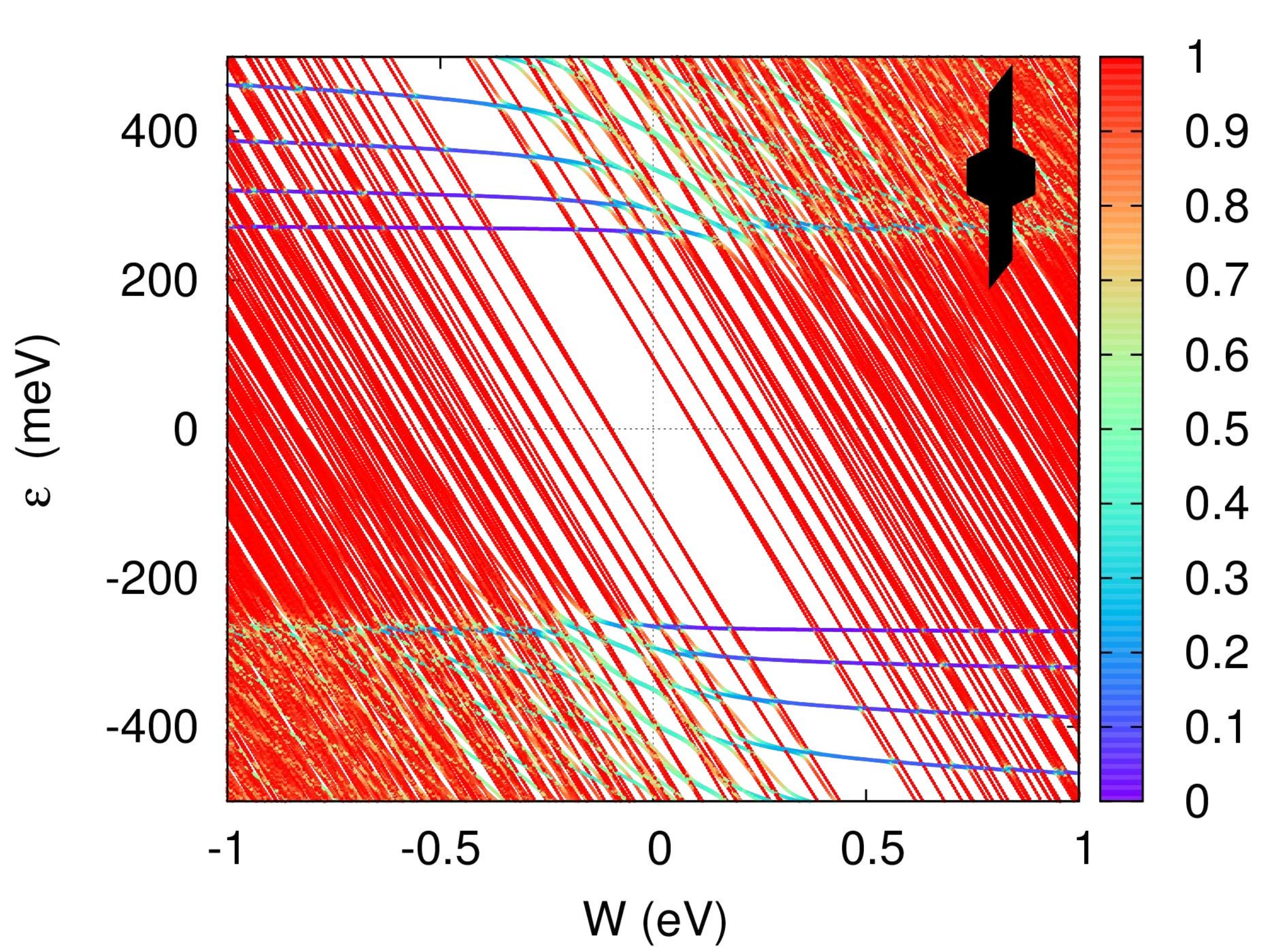} \\
 \end{tabular}
\caption{Single-electron spectra for a quantum dot of length $2R=20$ nm defined within a semiconducting armchair nanoribbon of length $L=60.1$ nm
and $18$ (a),
or $36$ (b) atoms across the channel. The colors of the lines
indicate the extent of the electron localization within the dot defined as the probability to find the electron within the segment $[-1.2R,1.2R]$ (see the color scale at right-hand side of the plots).
In (c) the results for the hexagonal flake inside the nanoribbon with 18 atoms are given. The black insets symbolically denote the studied system.
  }\label{ribo}
\end{center}
\end{figure}

\section{Results and Discussion}
\subsection{Single-electron spectra}
Let us first characterize the single-electron spectrum in the absence of the electron-electron interaction
for the quantum dots induced within the ribbon.
The energy spectrum is displayed in Fig. \ref{ribo}
 as a function of the quantum dot potential for a nanoribbon of length $L=61$ nm
and 18 or 36 atoms across the ribbon -- Fig. \ref{ribo}(a) and (b), respectively.
 The colors of the plotted energy levels indicate the localization of the corresponding states.
  The red energy levels are entirely localized within the quantum dot area $[-1.2R,1.2R]$.
  For the quantum dot induced by the external potential in the homogenous ribbon the
localized energy levels appear for $W\neq 0$.
The dot-localized (red) energy levels for $W>0$ correspond to localized states of the conduction band (n-type quantum dot)
and for $W<0$ to the localized states of the valence band (p-type quantum dot).
The energy levels plotted in blue in Fig. \ref{ribo} are localized entirely outside the quantum dot, generally near the ends of the ribbon,
and they are insensitive to the dot potential.

At the charge neutrality point ($E=0$) [Fig. \ref{ribo}(a,b)] one observes an energy gap  between the dot-localized (red) energy levels
as a function of the dot potential $W$. For $E=0$ we find localized conductance (valence) band states appearing for increasing $W>0$
(decreasing $W<0$).  A gap is also observed as a function of the energy for a fixed $W$.
In the experimental conditions the transport gap is observed
as a function of both the dot potential (top gate potential) and the Fermi energy (back gate potential) \cite{xliu}. The latter
 corresponds to variation of the energy for a fixed $W$ value.

Figure \ref{ribo}(c) shows the spectra for the  hexagonal flake
embedded inside the ribbon with 18 atoms across the channel.
For the flake the number of localized
(red) energy levels  [Fig. \ref{ribo}(c)] within the gap is much larger
as compared to the nanoribbon quantum dot [cf. Fig. \ref{ribo}(a)],
and the localized energy levels appear already at $W=0$.
%In Fig. \ref{flejk}(a-b) we can see a number of horizontal
%energy levels that are insensitive to the dot potential,
%with the energy positions similar to the ones found
%for the ribbon quantum dot [Fig. \ref{ribo}(a)].

\begin{figure*}[htbp]
\begin{center}
\begin{tabular}{llll}
% a) & \includegraphics[angle=270,width=0.4\textwidth]{ChPot36.pdf}
% b)&\includegraphics[angle=270,width=0.4\textwidth]{Charge36.pdf}\\
%
% c) & \includegraphics[angle=270,width=0.4\textwidth]{ChPot18.pdf}
% d)&\includegraphics[angle=270,width=0.4\textwidth]{Charge18.pdf} \\
%
% e) & \includegraphics[angle=270,width=0.4\textwidth]{ChPot18Hex44.pdf}
% f)&\includegraphics[angle=270,width=0.4\textwidth]{Charge18Hex44.pdf} \\
%
% g) & \includegraphics[angle=0,width=0.4\textwidth]{ChPot18Hex44dluzszy.png}
% h) &\includegraphics[angle=0,width=0.4\textwidth]{Charge18Hex44dluzszy.png} \\

a) & \includegraphics[angle=0,width=0.44\textwidth]{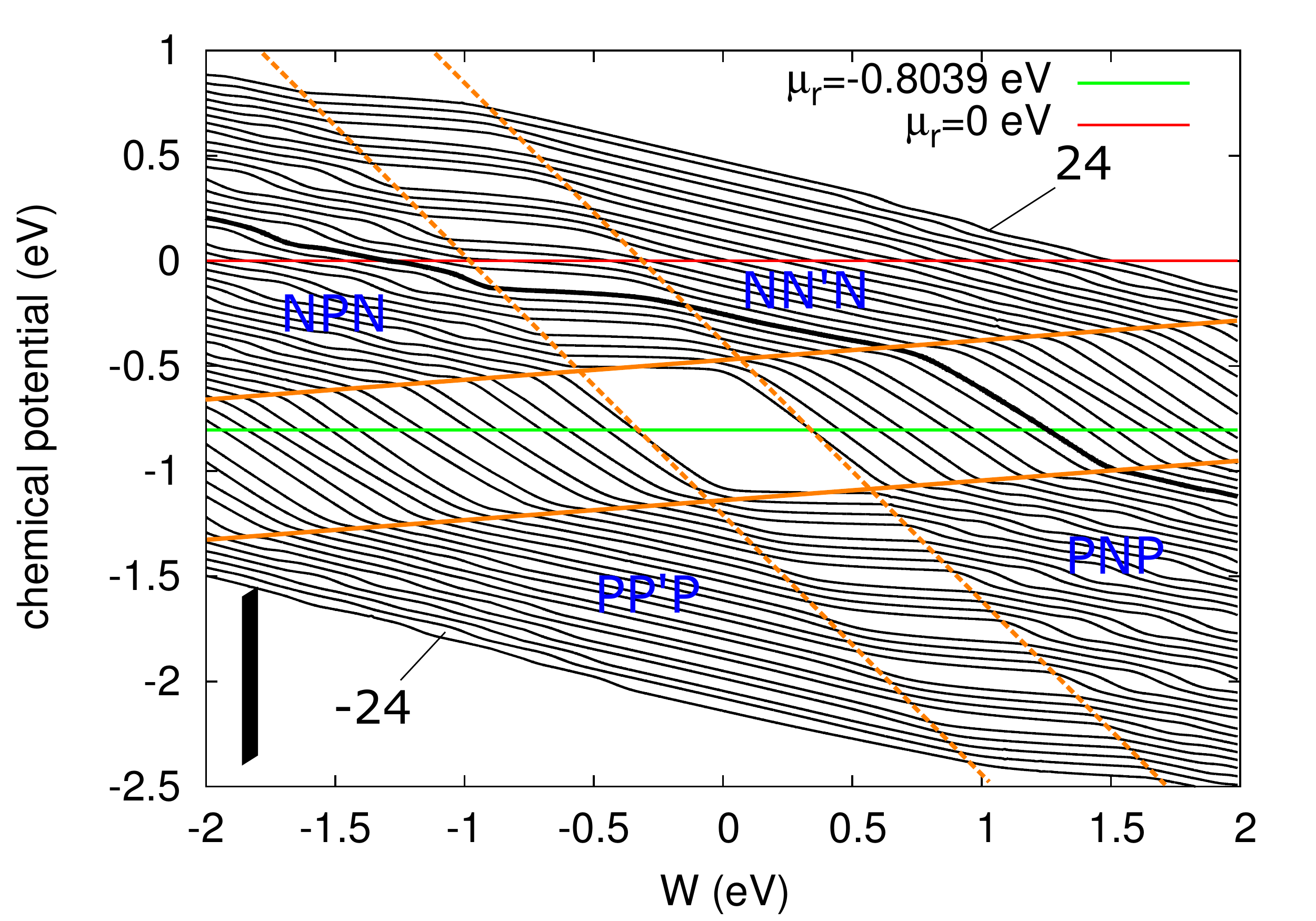}
b)&\includegraphics[angle=0,width=0.44\textwidth]{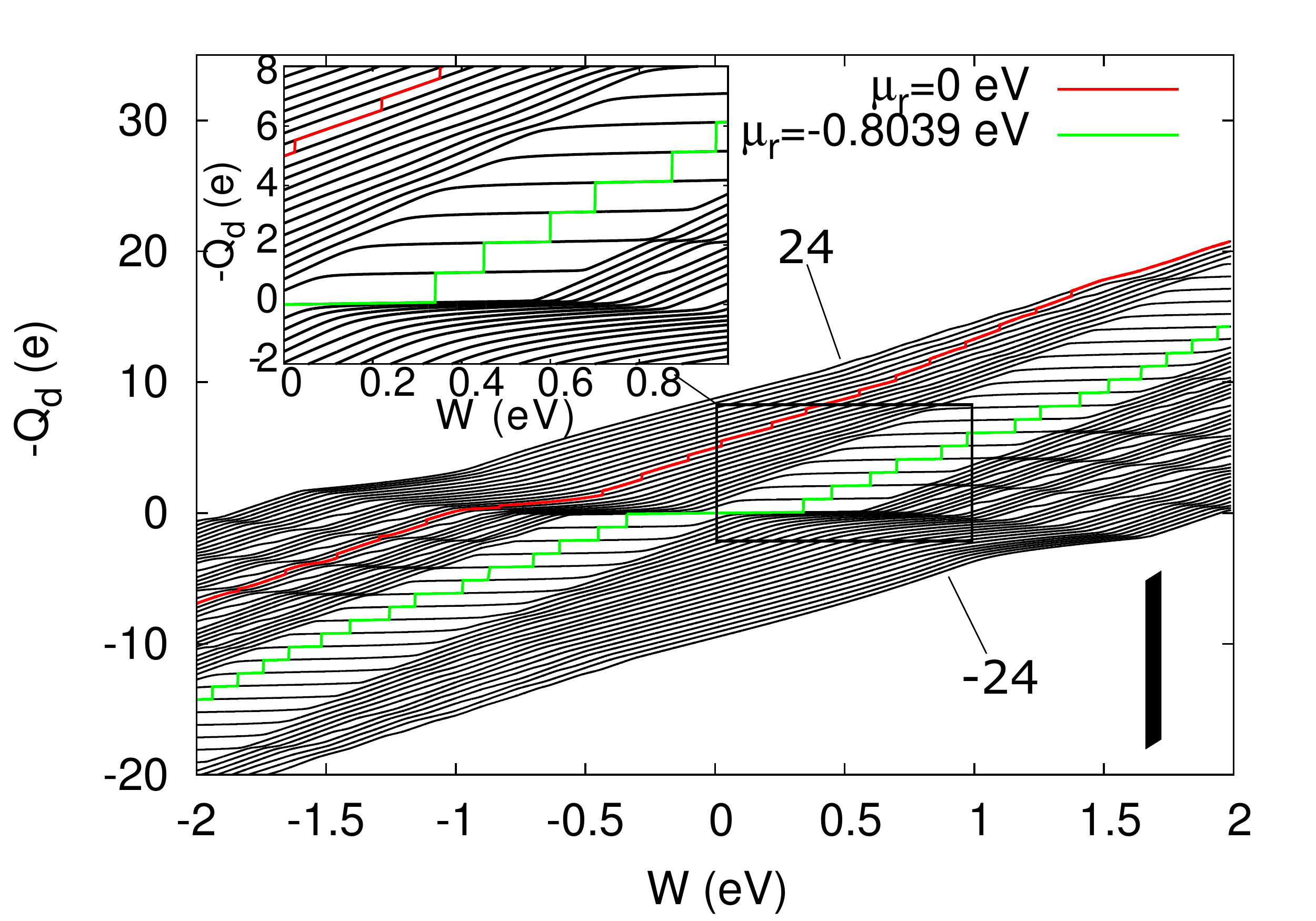}\\

c) & \includegraphics[angle=0,width=0.44\textwidth]{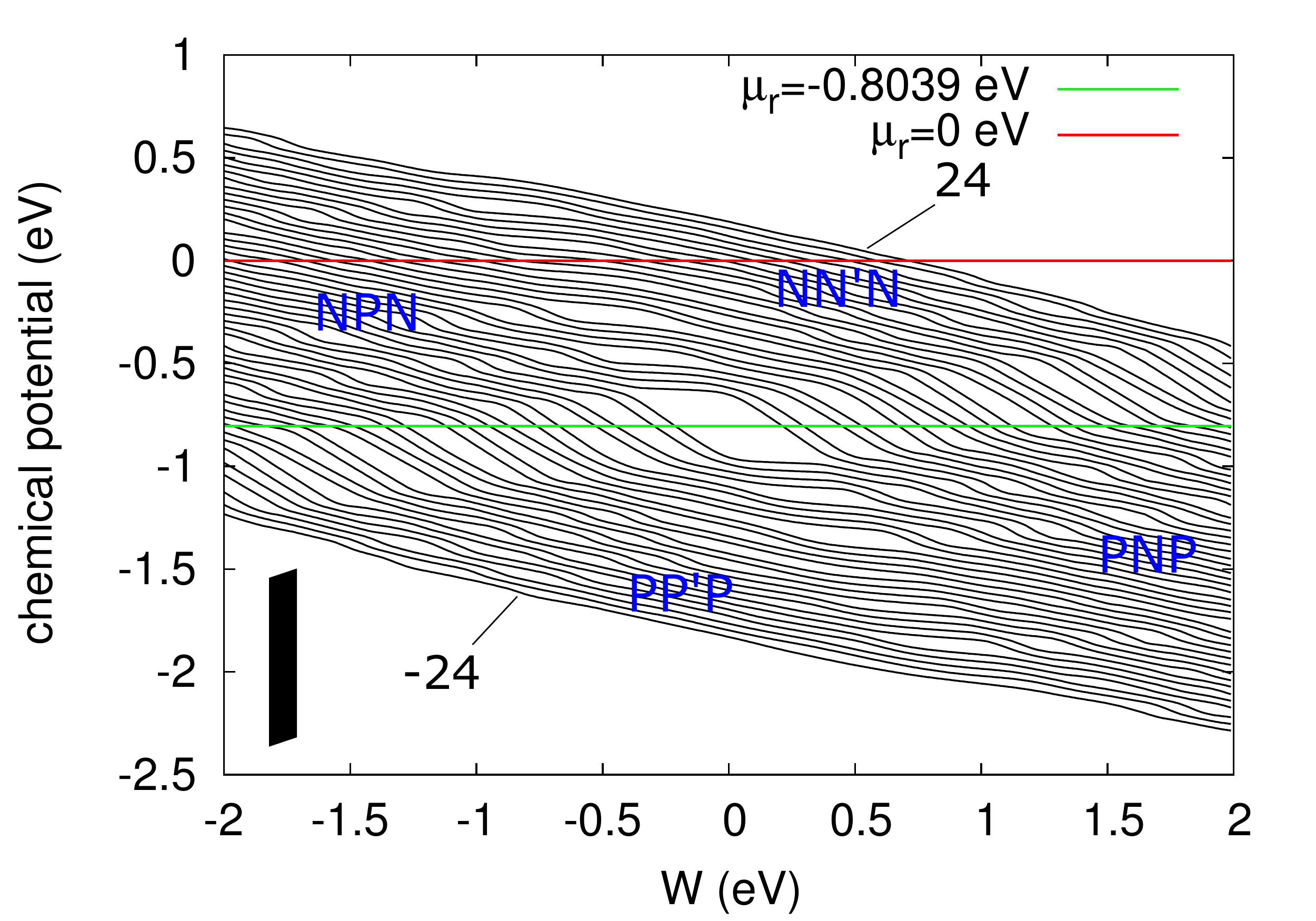}
d)&\includegraphics[angle=0,width=0.44\textwidth]{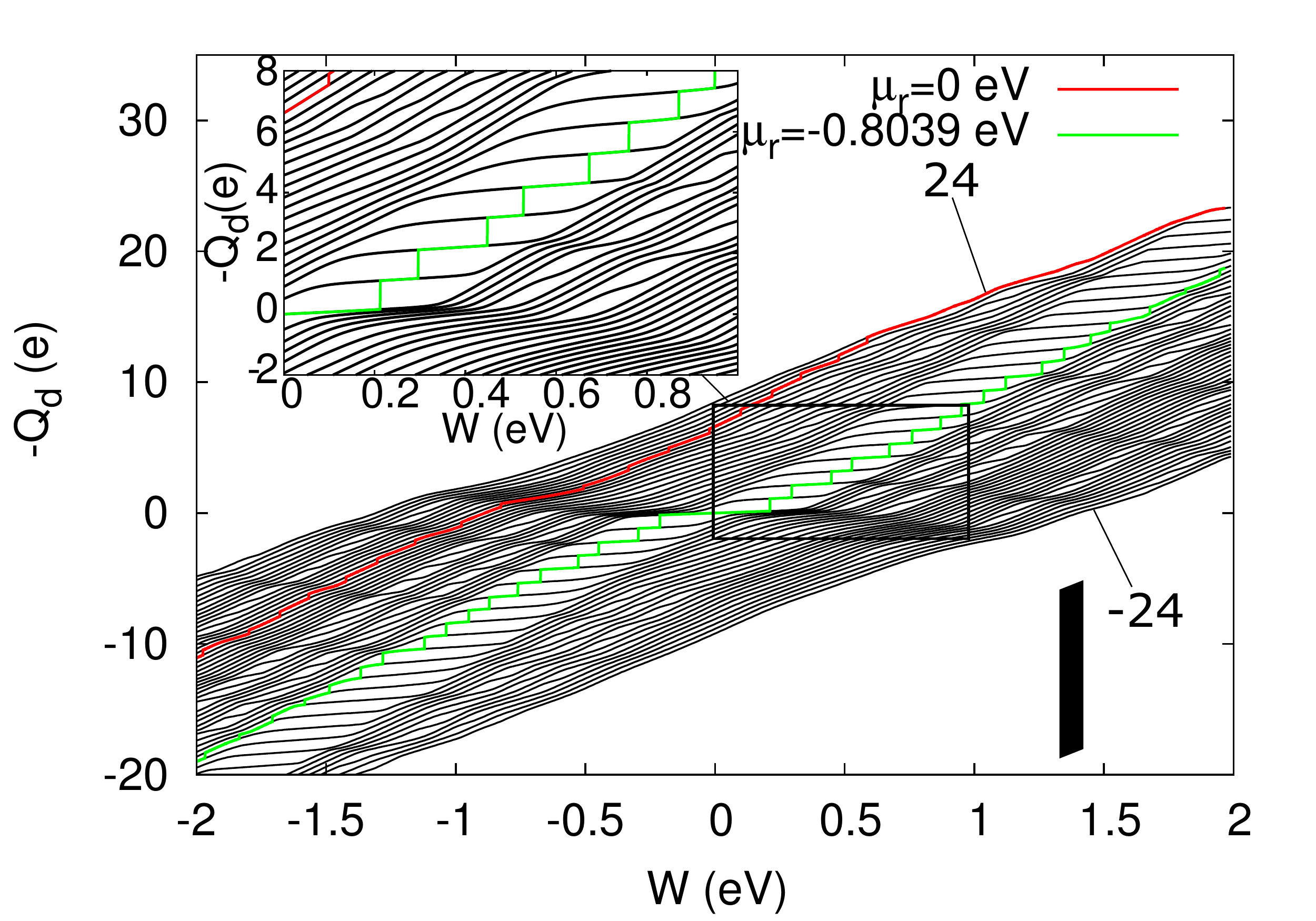} \\

e) & \includegraphics[angle=0,width=0.44\textwidth]{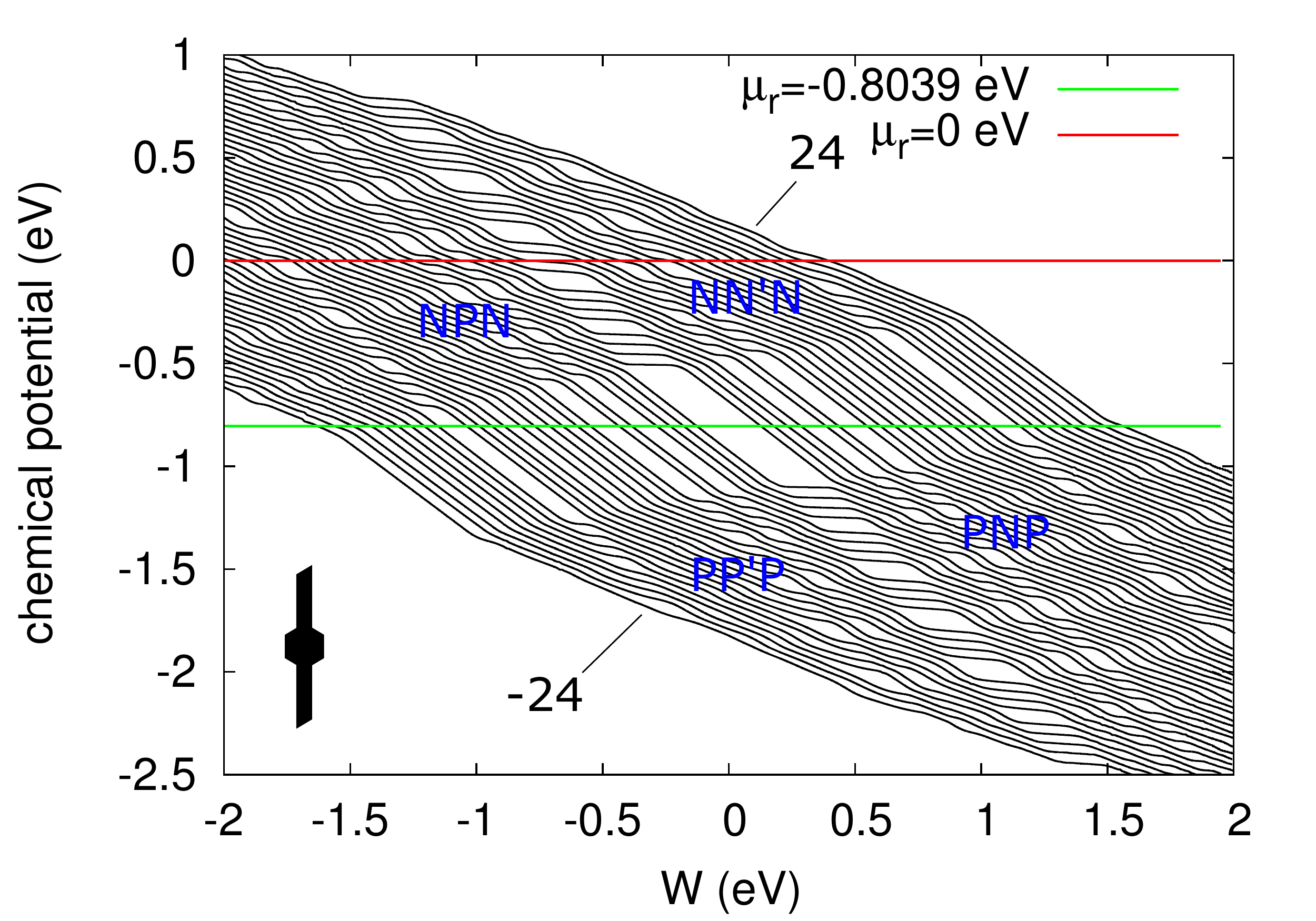}
f)&\includegraphics[angle=0,width=0.44\textwidth]{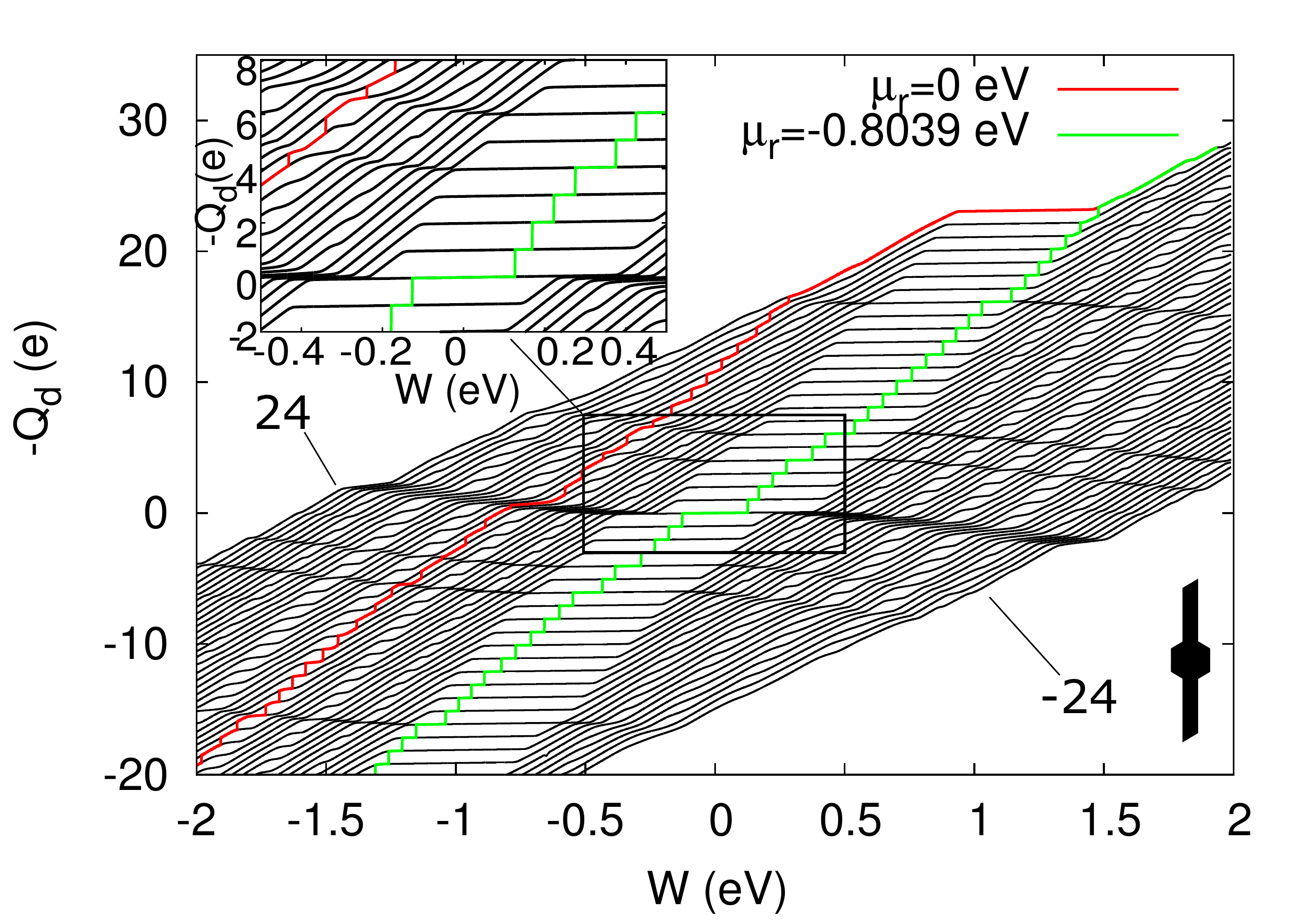} \\

g) & \includegraphics[angle=0,width=0.44\textwidth]{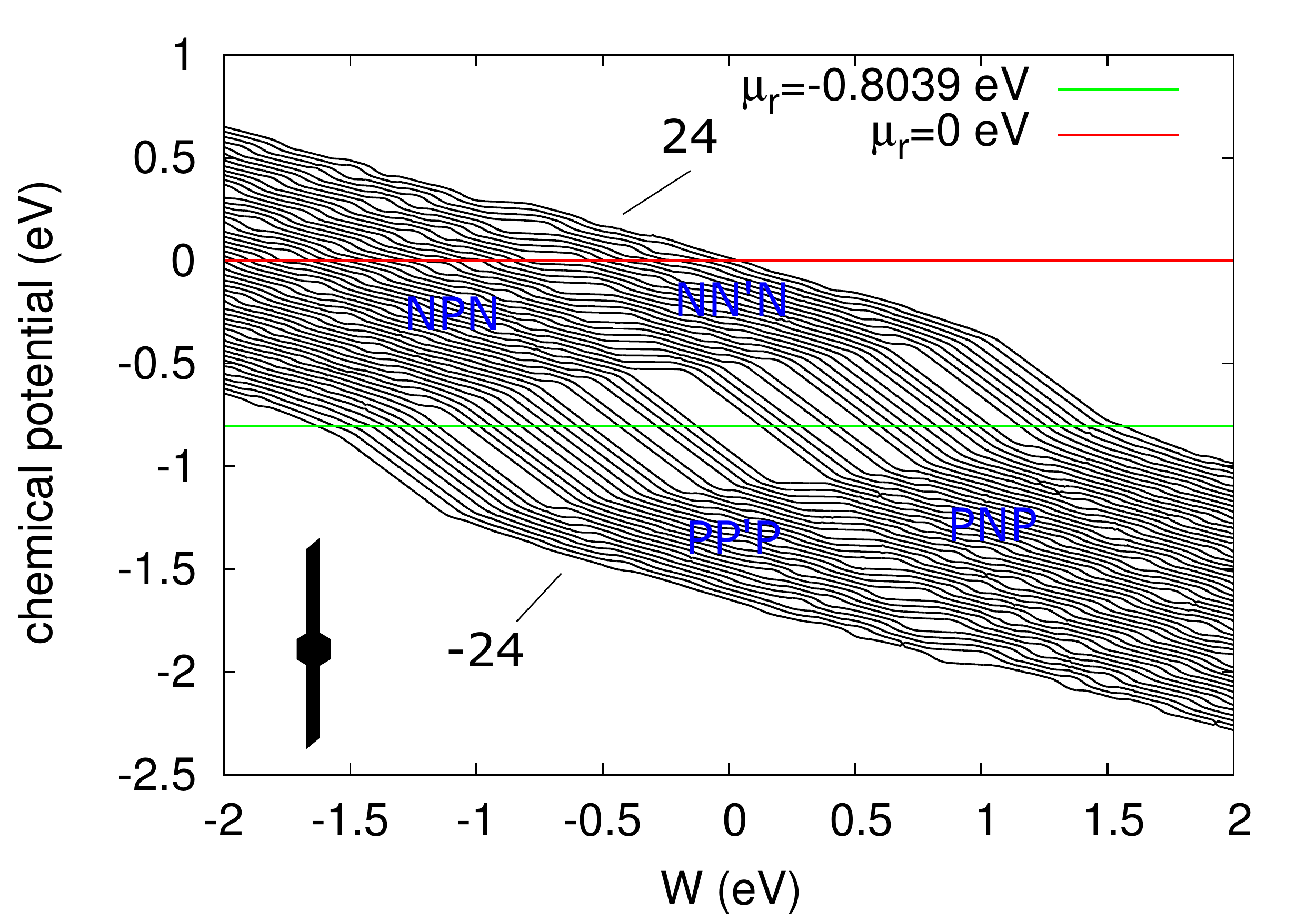}
h) &\includegraphics[angle=0,width=0.44\textwidth]{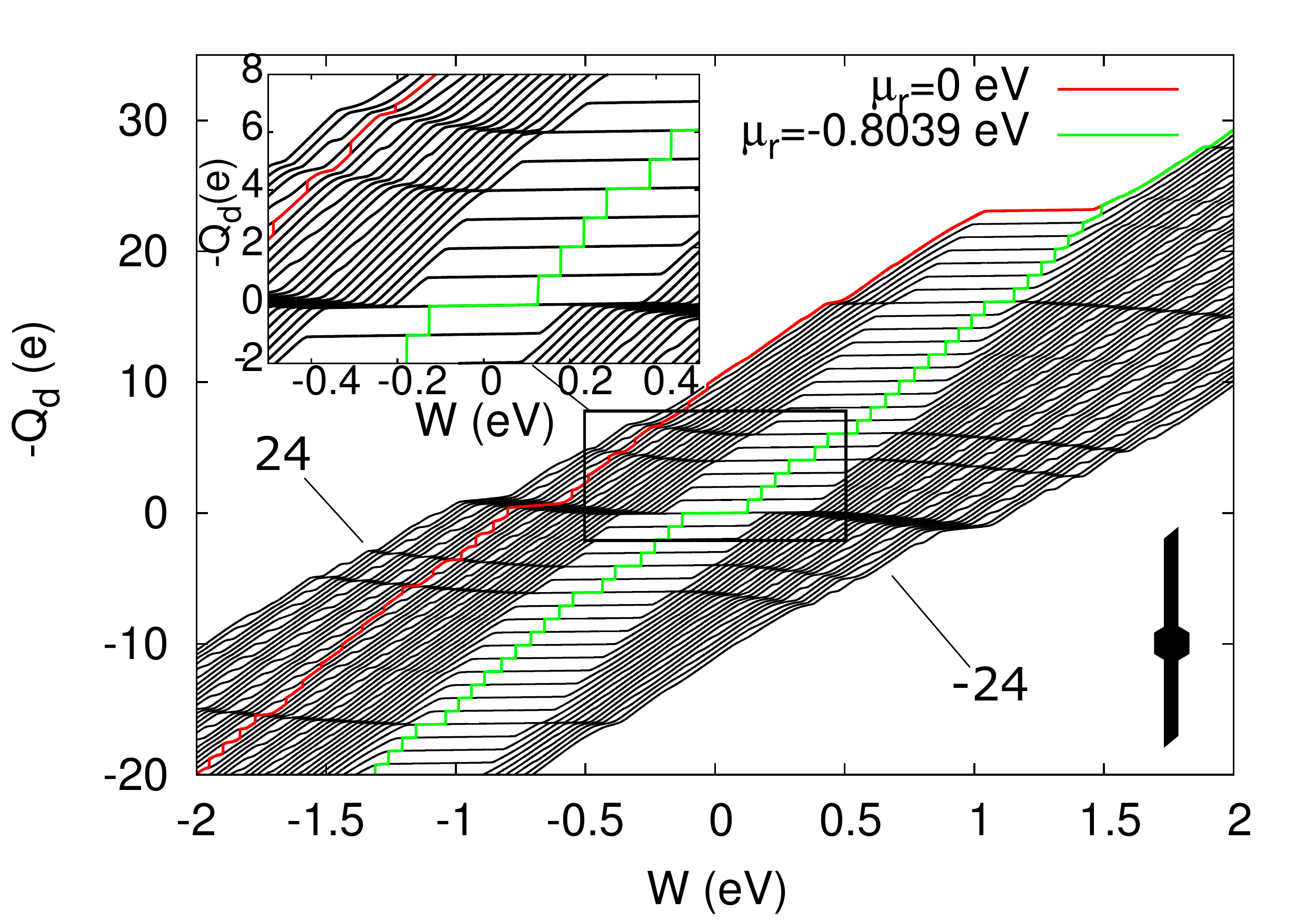} \\

\end{tabular}
\caption{(a) Chemical potentials for $N_e=-24,\dots 24$ excess electrons
in a 18-atoms wide nanoribbon as a function of the QD potential $W$.
(b) Charge contained within the quantum dot (segment $[-1.2R,1.2R]$) defined within the nanoribbon 18-atoms wide
as a function of $W$ for $N_e=-24,\dots 24$ excess electrons.
(c,d) -- same as (a,b) for nanoribbon with 36 atoms across the channel.
(e,f)-- same as (c,d) but for the hexagonal flake with 44 atoms along the side of the hexagon.
In (a-f) and elsewhere in the paper the length of the system is $L=60.1$ nm.
(g,h) - same as (g,h) but with longer ribbons with $L$ increased to 90 nm.
Insets in the figures shows an enlarged fragment from the graph.
The orange lines in (a) mark the transport gap due to the Fermi energy (solid lines) and (b) the external potential (dotted lines) -- see text.
The thicker line in (a) shows the chemical potential of 8 electron confined in the dots $\mu_8$ -- which is discussed in detail in the text.
 }\label{flejkcz}
\end{center}
\end{figure*}

\begin{figure}[htbp]
\includegraphics[scale=0.2]{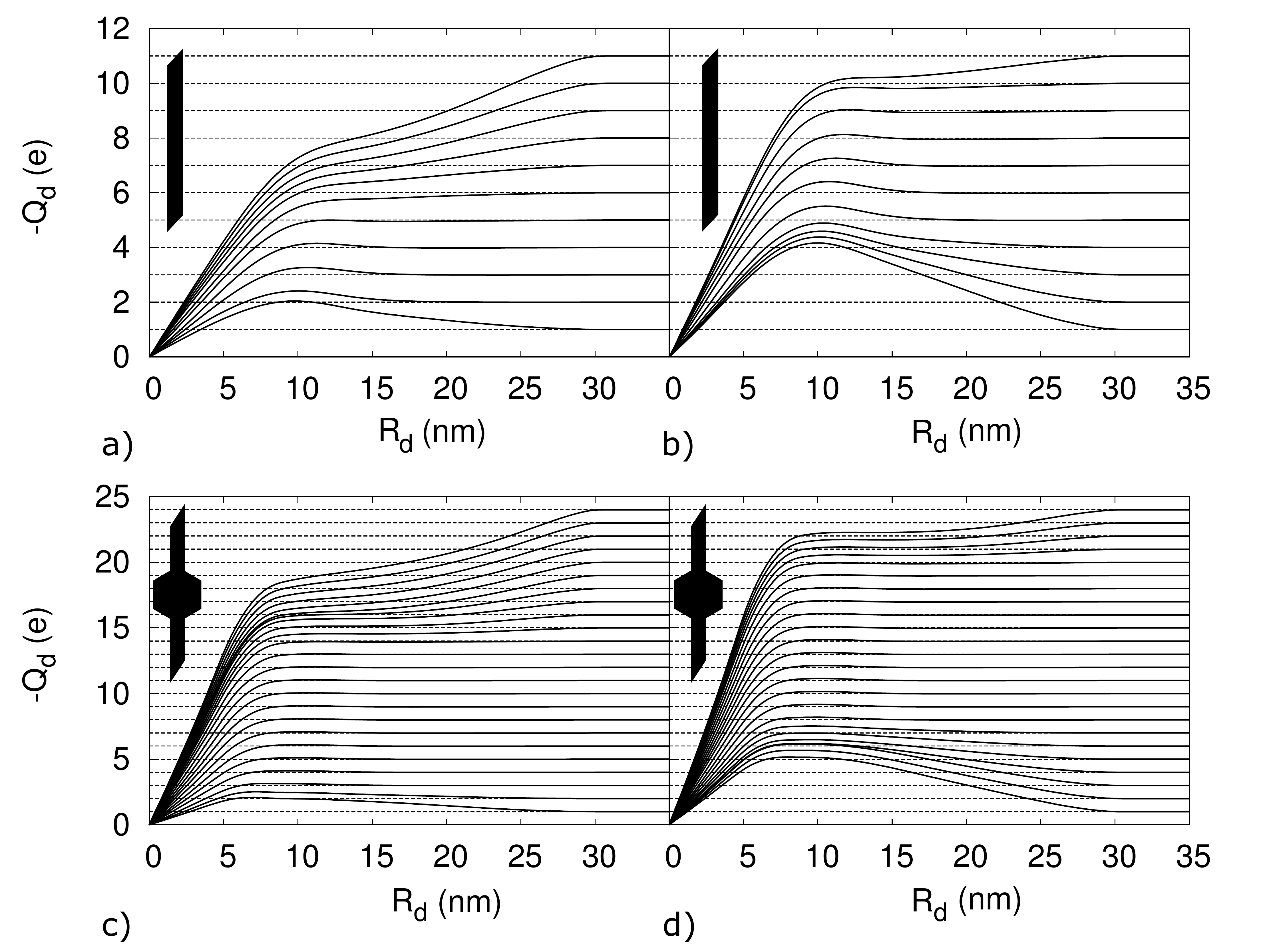}
% \begin{tabular}{llll}
% a) & \includegraphics[angle=0,width=0.23\textwidth]{Qd36W=500.pdf}  &
% b) & \includegraphics[angle=0,width=0.23\textwidth]{Qd36W=800.pdf} \\
% c) & \includegraphics[angle=0,width=0.23\textwidth]{Qd18Hex44W=500.pdf} &
% d) & \includegraphics[angle=0,width=0.23\textwidth]{Qd18Hex44W=800.pdf}
% \end{tabular}
\caption{ (a,b) Charge contained within a segment of $[-1.2R_d,1.2R_d]$
for $N_e=1,2,\dots$ excess electrons for the quantum dot defined within the 36-atoms wide ribbon width for $W=0.5 $ eV (a) and $W=0.8$ eV.
For the energy spectrum and the charge contained within the quantum dot see Fig. \ref{flejkcz}(a,b) for $W=0.5$ eV (a) and $W=0.8$ eV (b).
(c,d) -- same as (a,b) only for a semiconducting 18-atoms ribbon containing a hexagonal flake  across the channel.}
\label{qdrd}
\end{figure}

\begin{figure*}[ht!]
% \begin{center}
\begin{tabular}{ccc}
a) \includegraphics[angle=0,width=0.2\textwidth]{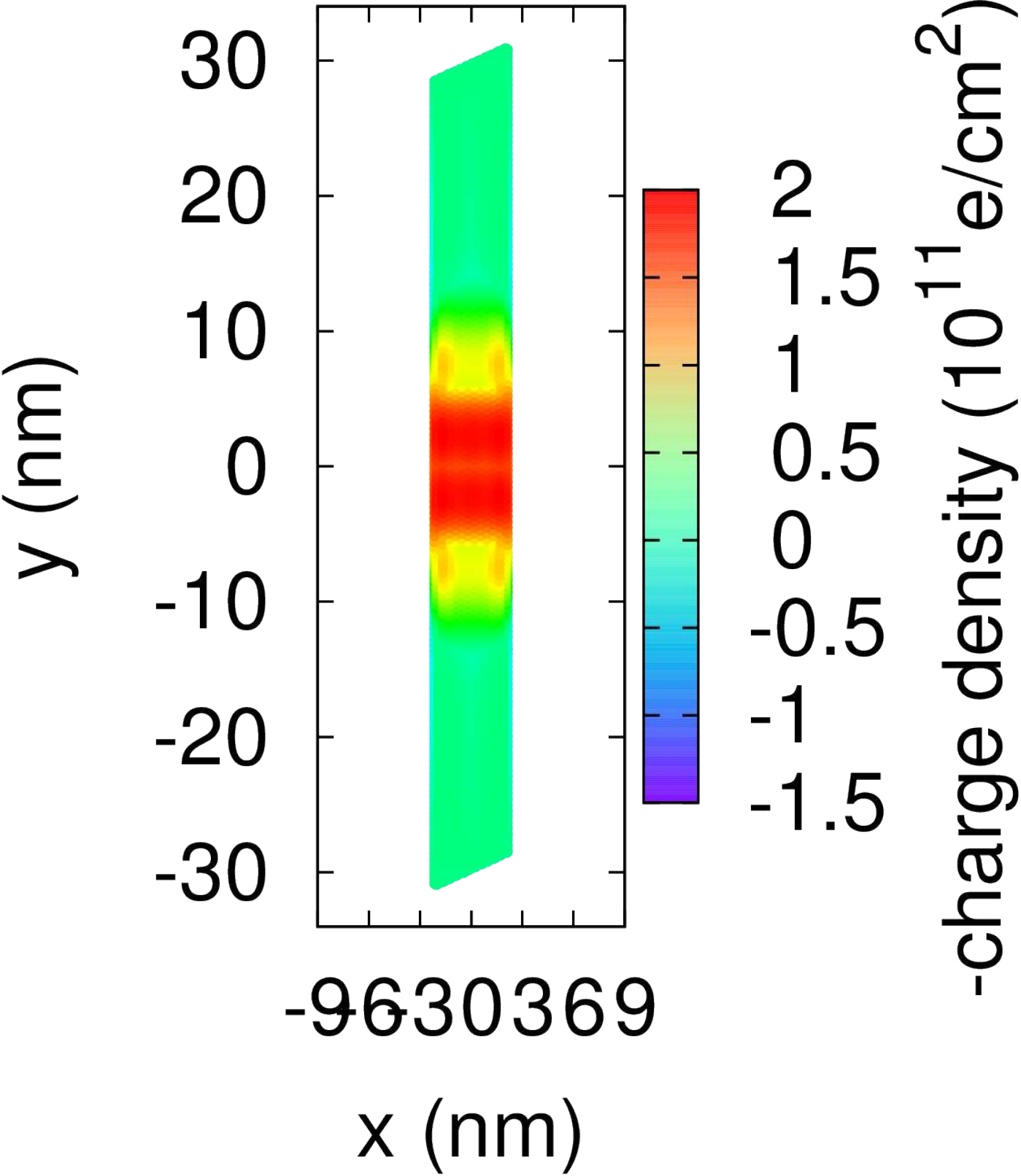}
b) \includegraphics[angle=0,width=0.21\textwidth]{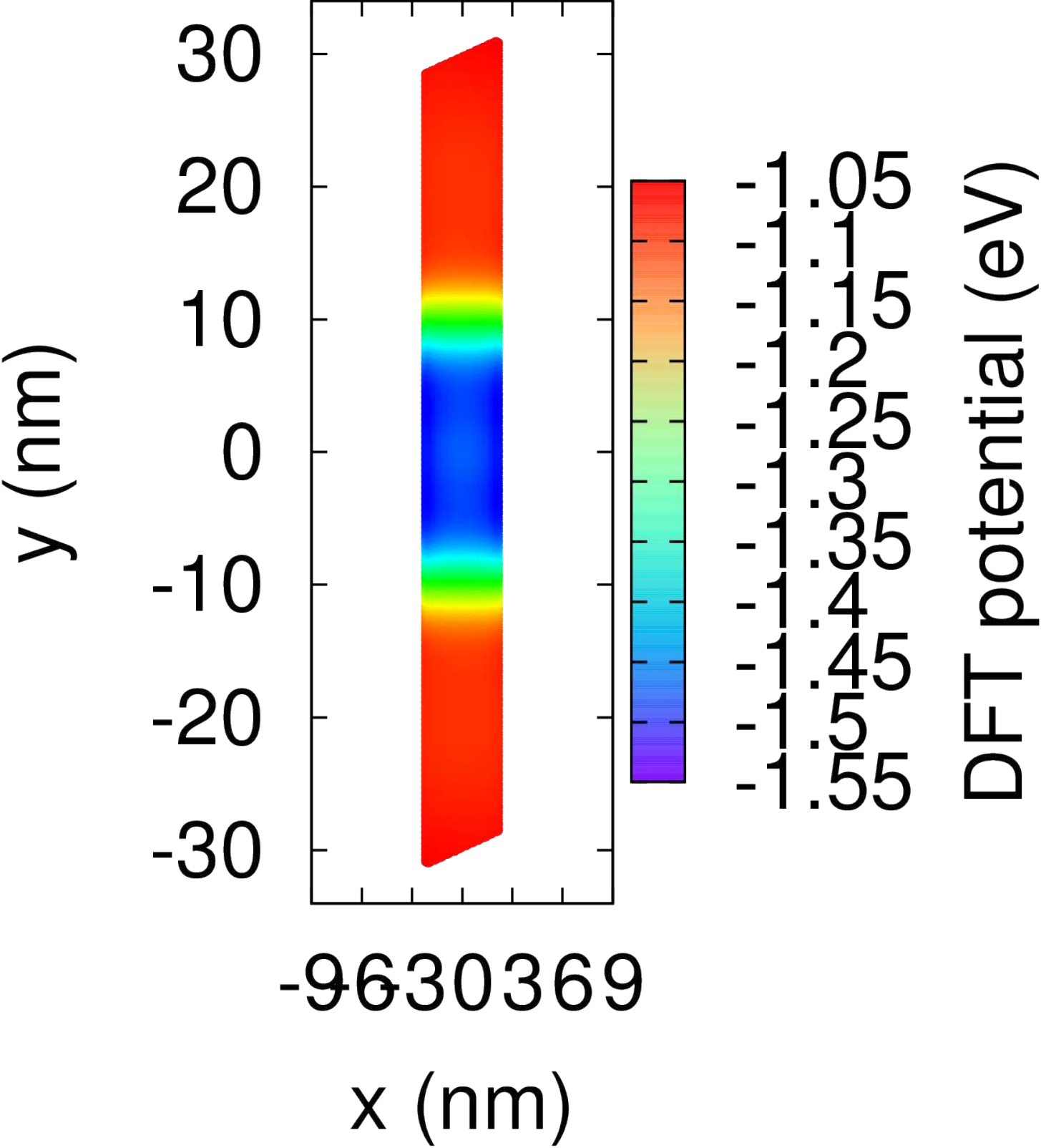}
c) \includegraphics[angle=0,width=0.435\textwidth]{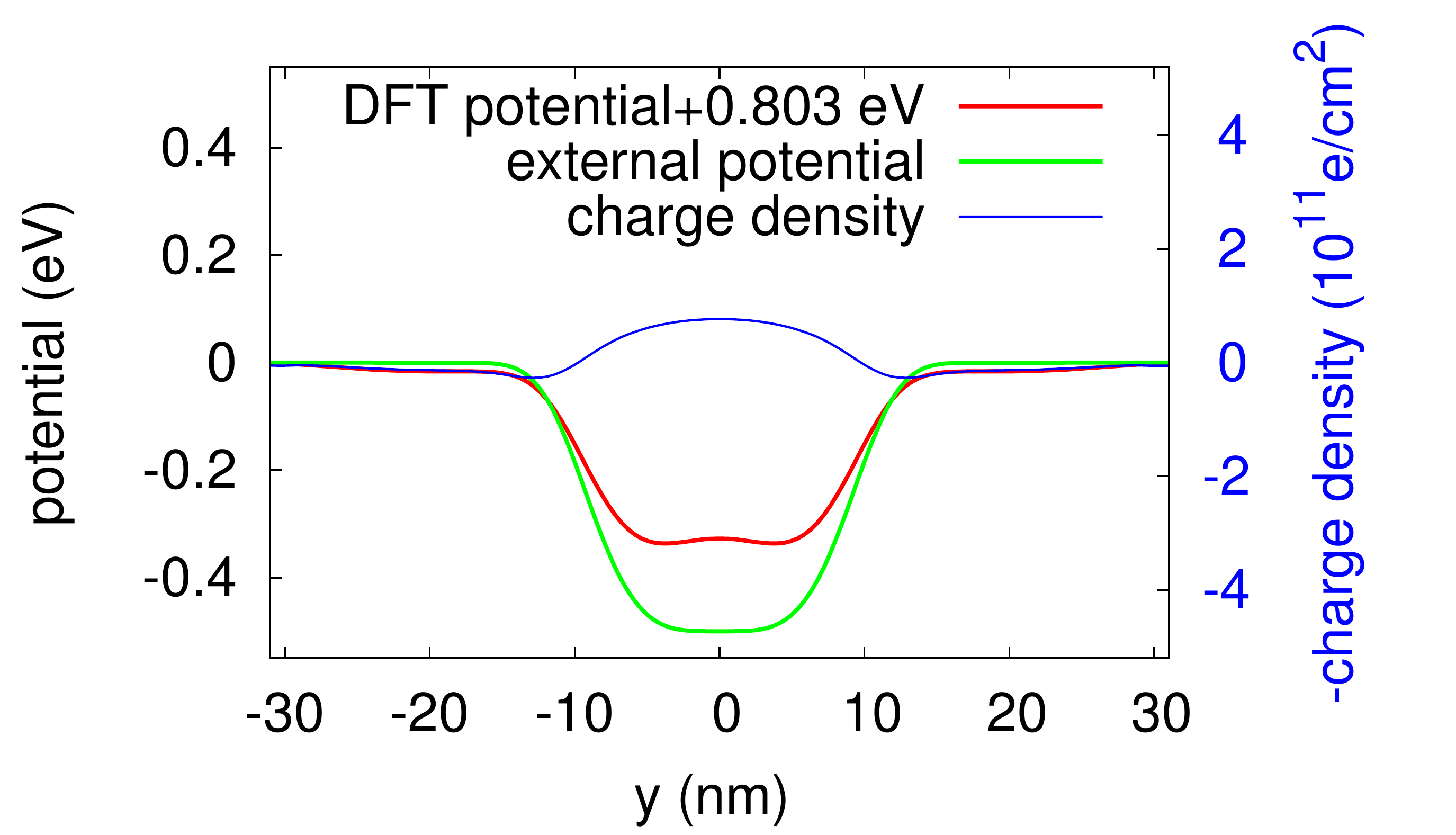}\\

d) \includegraphics[angle=0,width=0.2\textwidth]{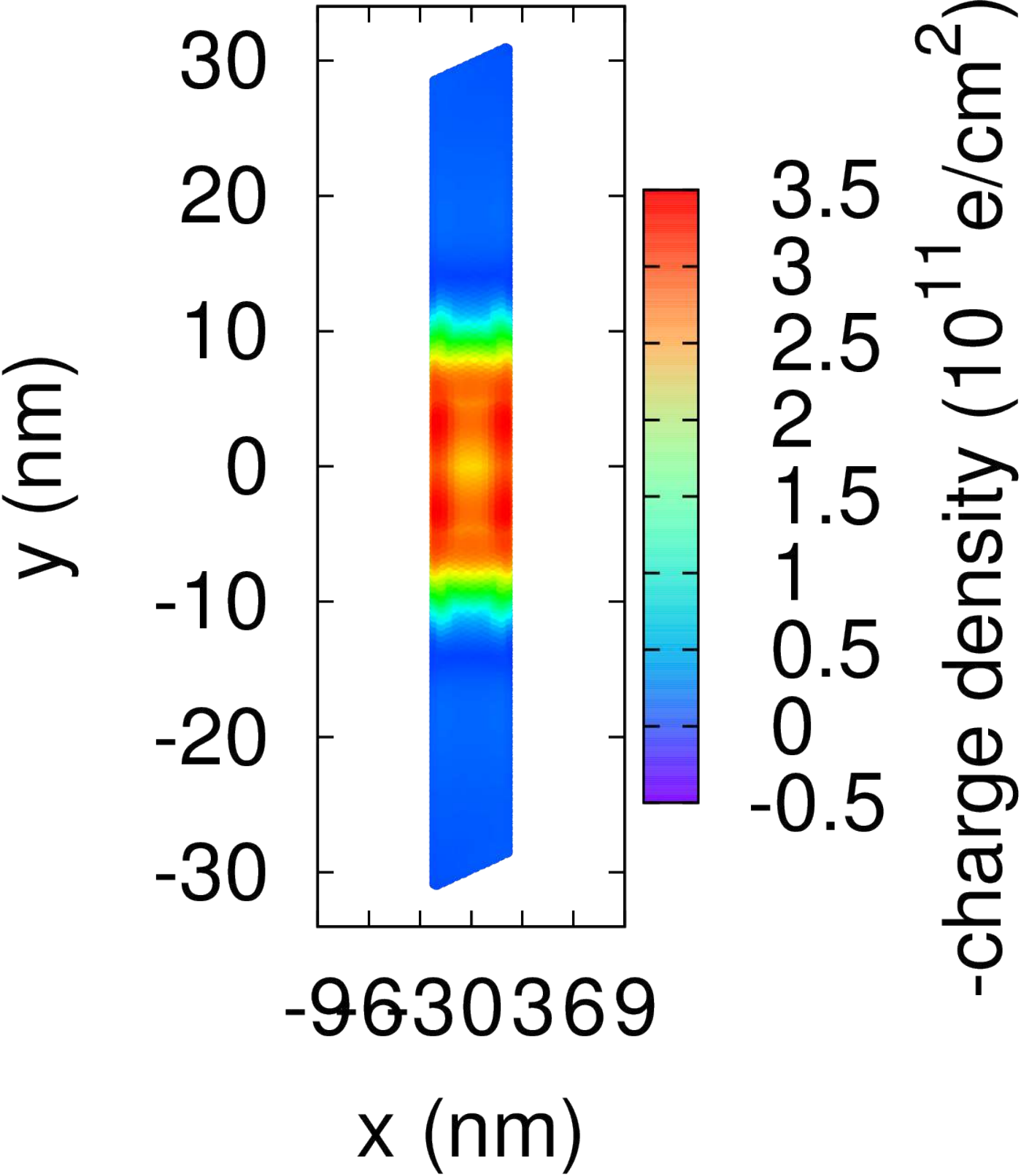}
e) \includegraphics[angle=0,width=0.21\textwidth]{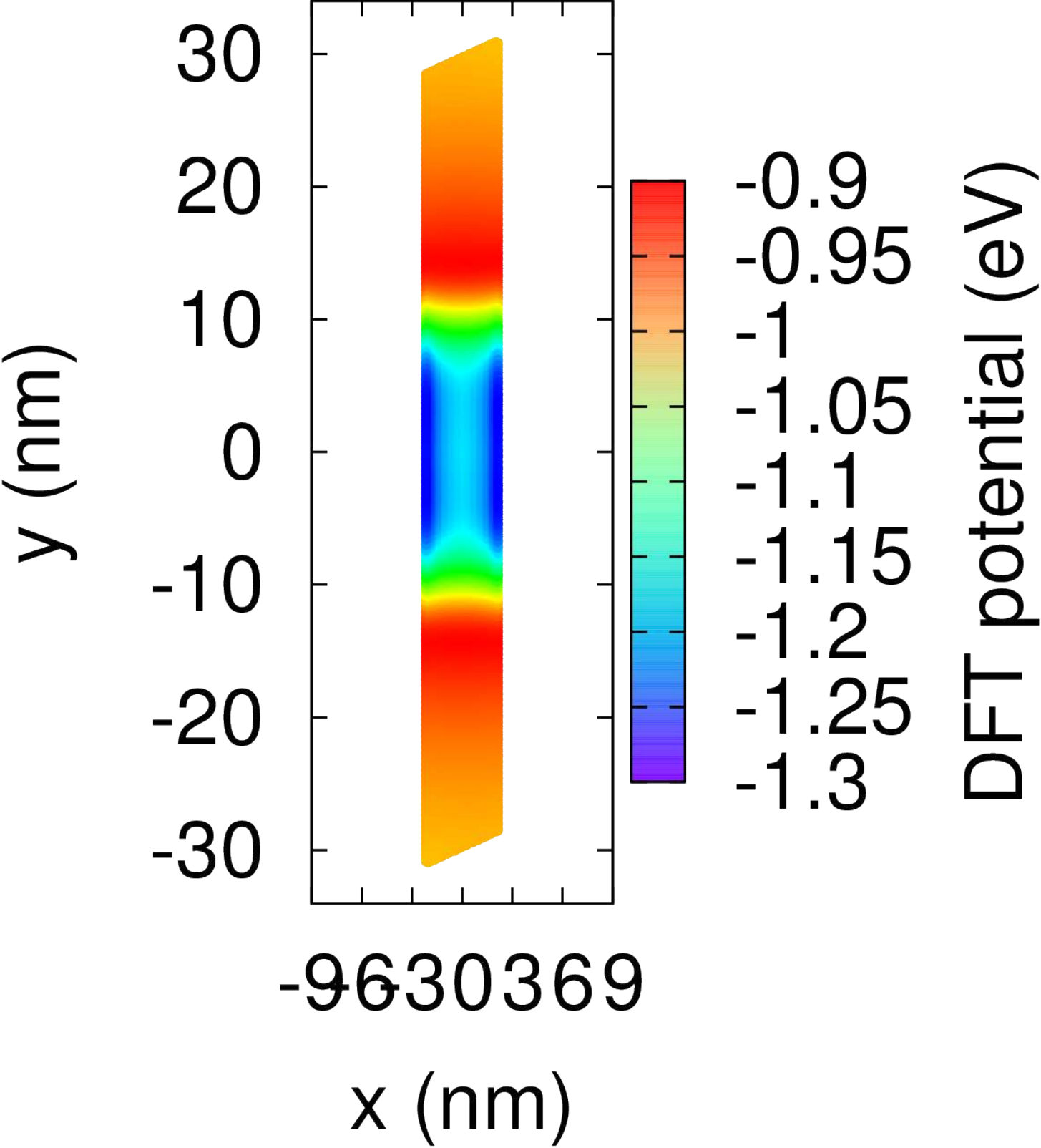}
f) \includegraphics[angle=0,width=0.435\textwidth]{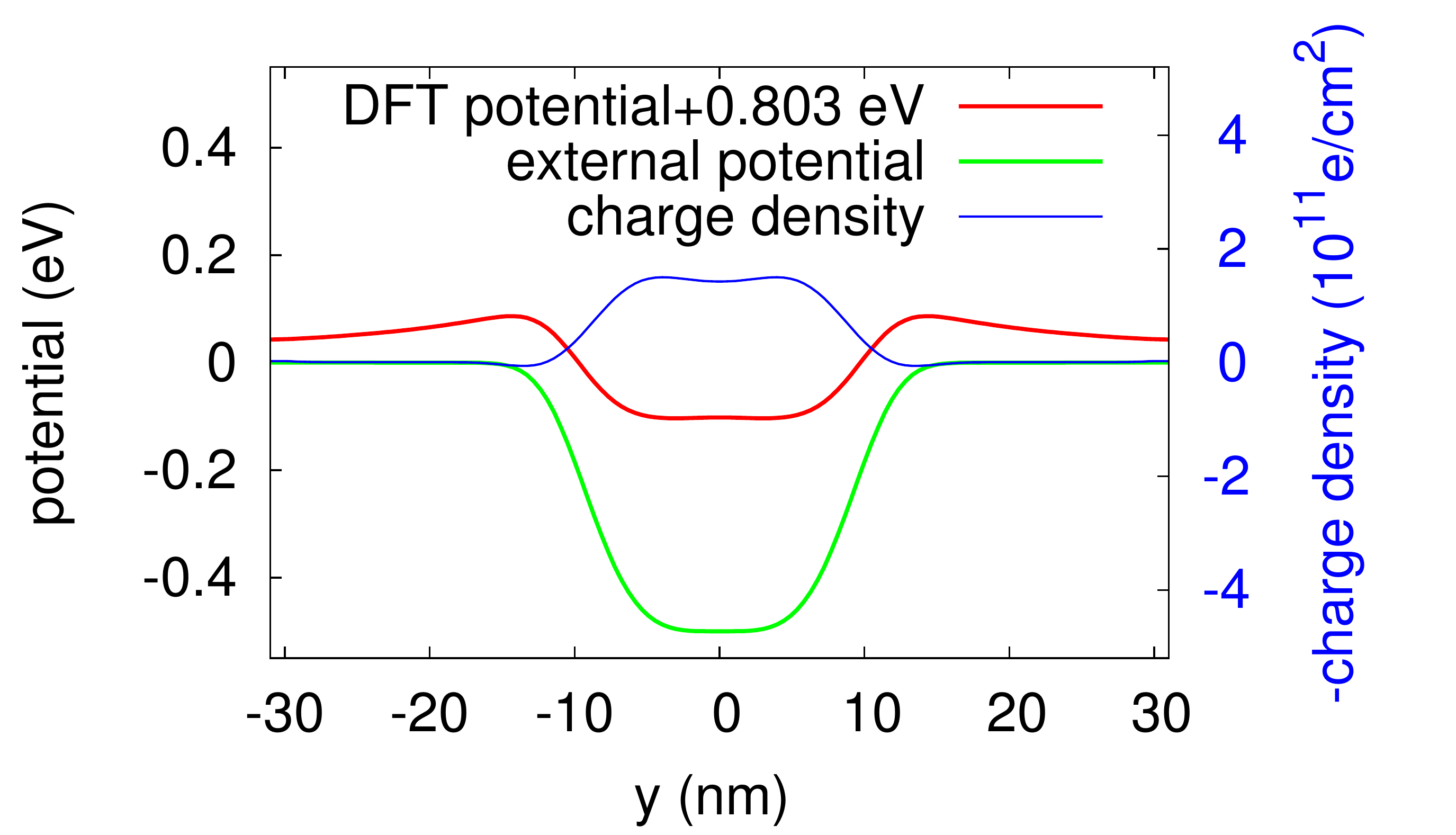}\\

g) \includegraphics[angle=0,width=0.2\textwidth]{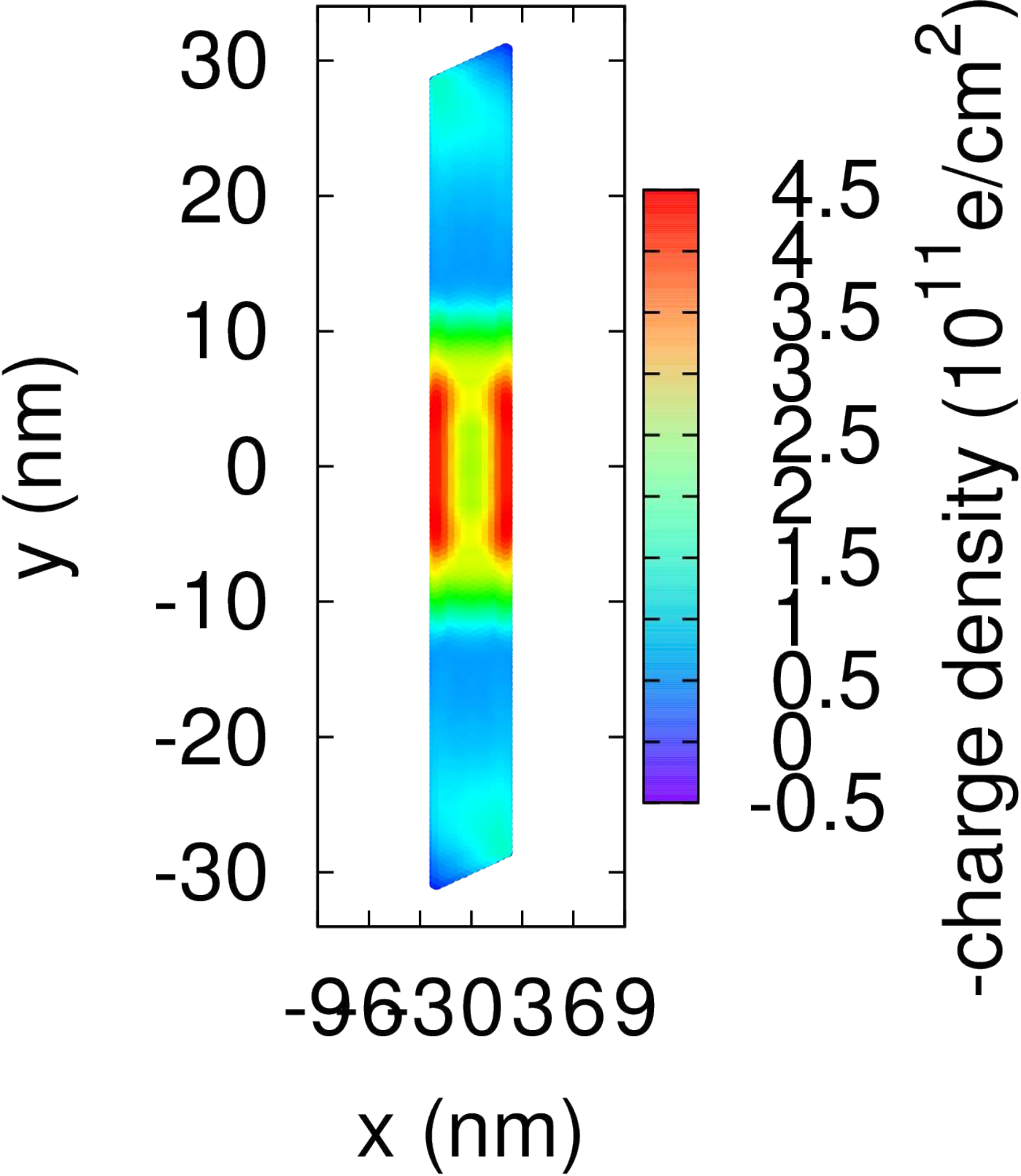}
h) \includegraphics[angle=0,width=0.21\textwidth]{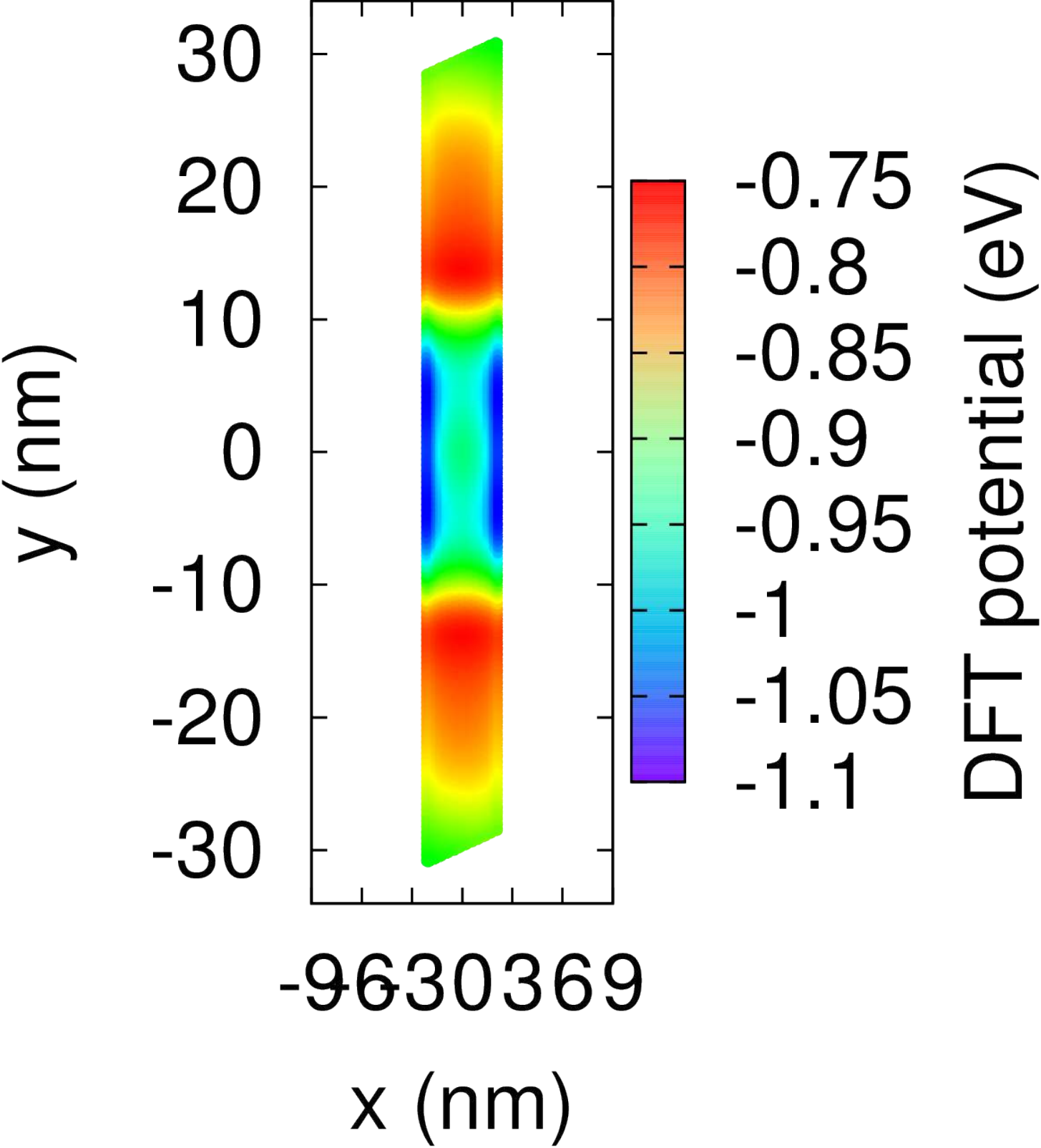}
i) \includegraphics[angle=0,width=0.435\textwidth]{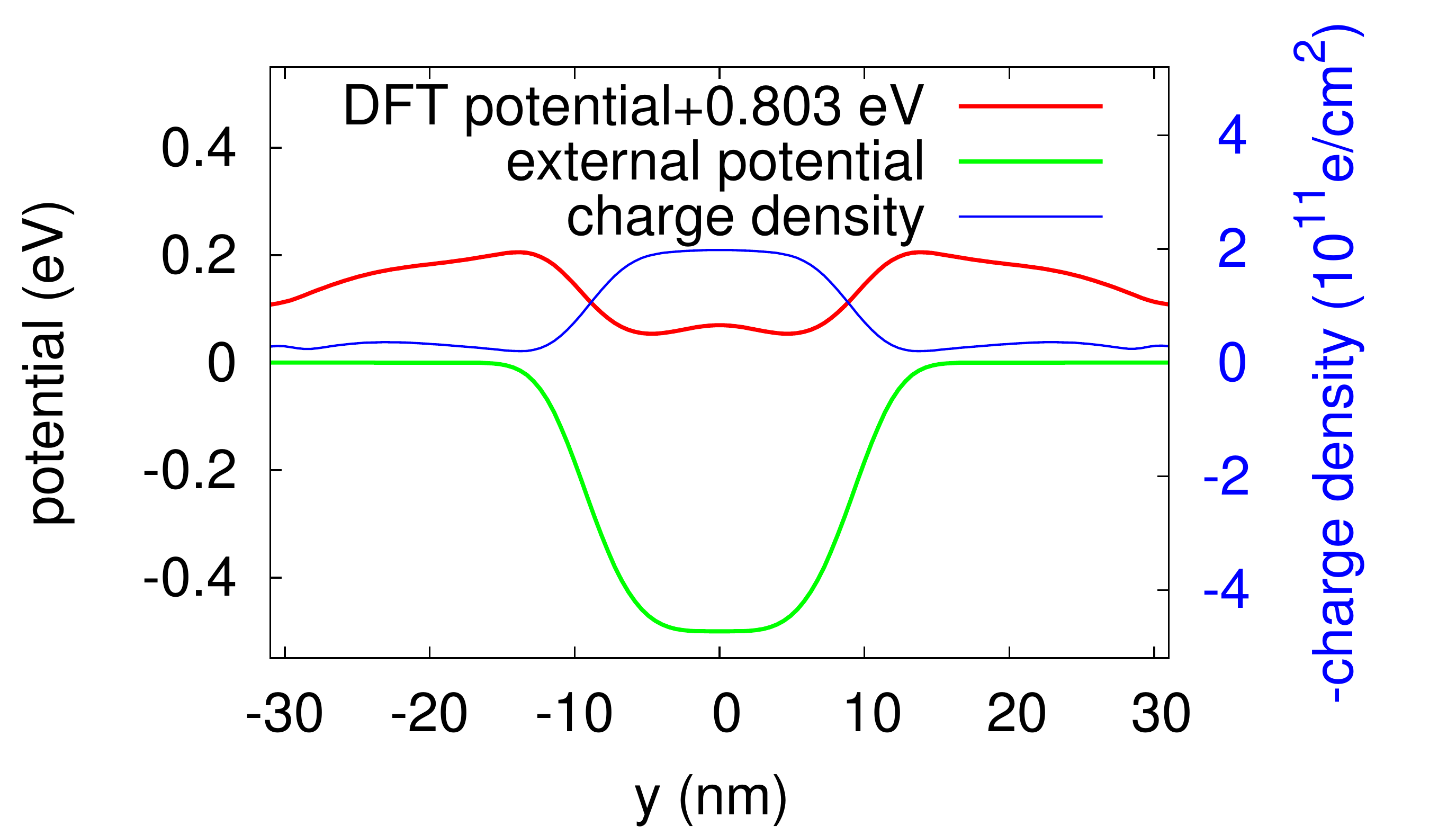}\\
 \end{tabular}
\caption{Excess electron charge (a,d,g) and the total potential (b,e,h) for the nanoribbon
of Fig. \ref{flejkcz}(a,b) at $W=0.5$ eV for $N_e=1,5$ and 9 respectively.
(c,f,i) present the cross sections of a-b, d-e, g-h, along the axis of the ribbon,
respectively, compared the external potential of Eq. (\ref{Vext}).}
\label{pcjaly}
% \end{center}
\end{figure*}

\begin{figure}[htbp]
\begin{tabular}{ll}
a) & \includegraphics[angle=0,width=0.45\textwidth]{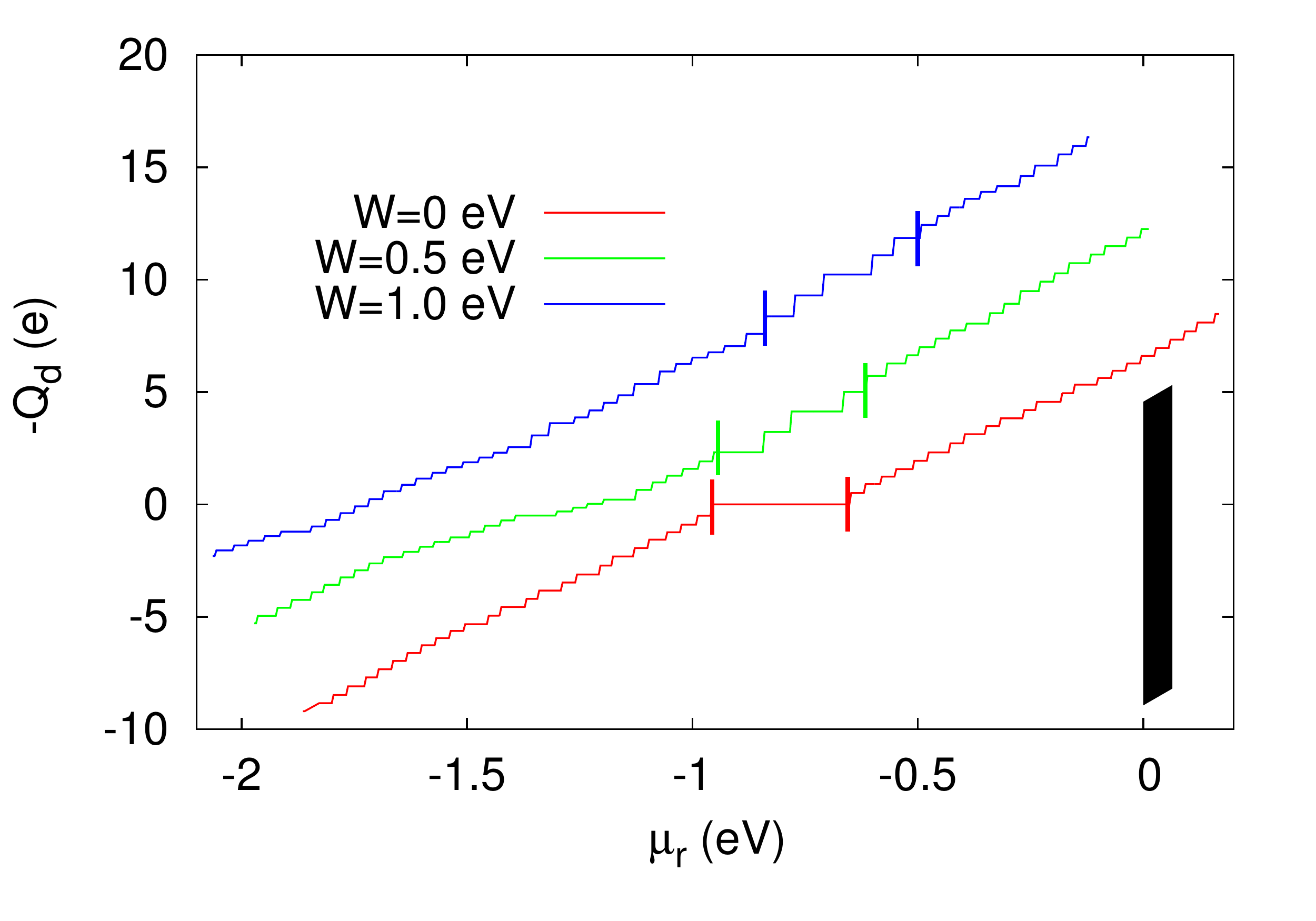}\\
b) & \includegraphics[angle=0,width=0.45\textwidth]{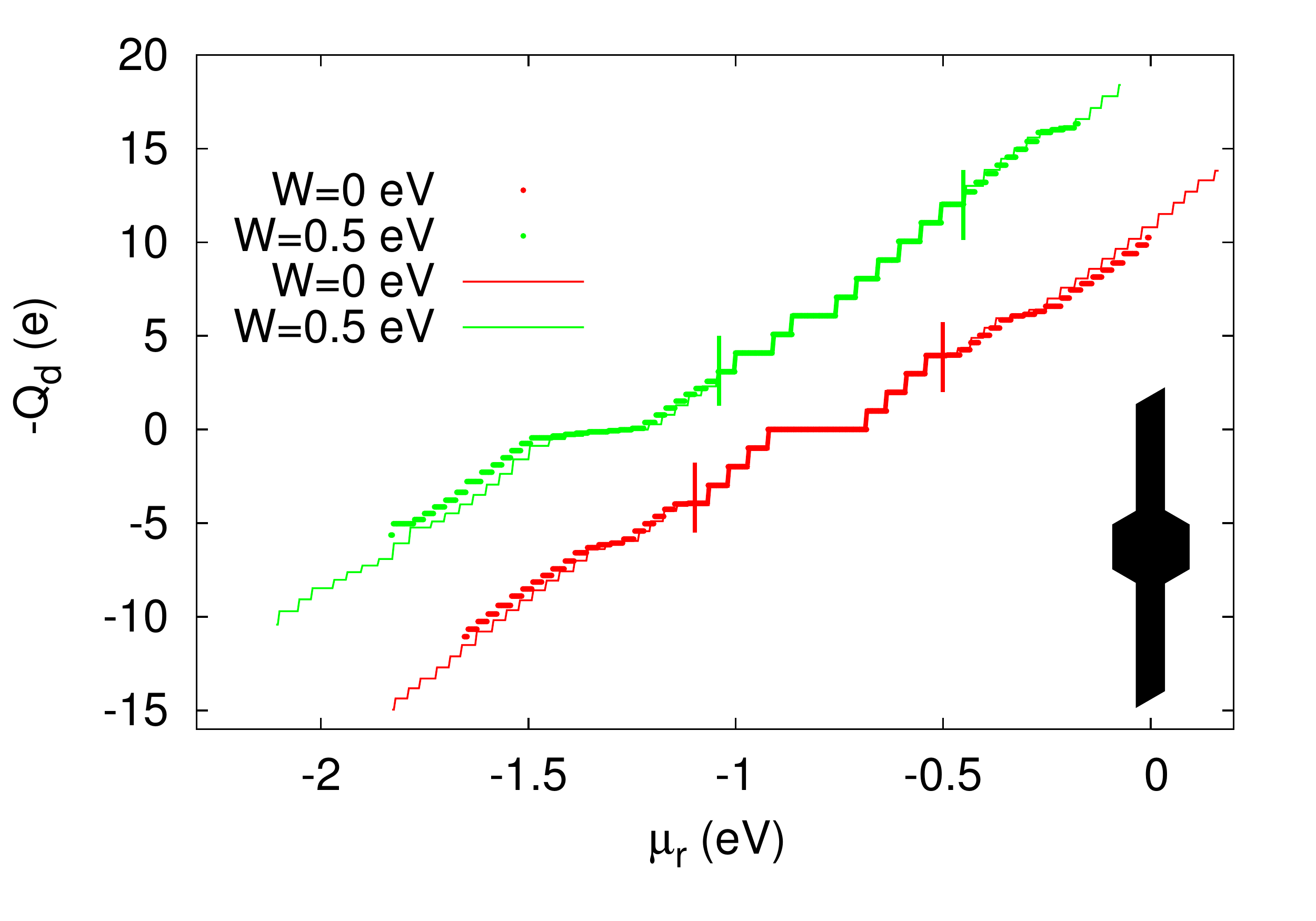}
\end{tabular}
\caption{The charge localized in the quantum dot $Q_d$ of a 36-atoms-wide nanoribbon (a) and
the hexagonal flake embedded within 18-atoms-wide ribbon (b) for $W=0$, 0.5 and 1 eV
as a function of the electrochemical potential of the external electron reservoir $\mu_r$.
The region of the transport gap is marked with the vertical lines. In (b) with the lines (dots) we marked the results
obtained for nanoribbon length $L=60.1$ nm  (90 nm). For the chemical potential spectra see Fig. \ref{flejkcz}(a) and Fig. \ref{flejkcz}(e,g), for the system without (with)
the flake inside the nanoribbon, respectively.}
 \label{wt}
\end{figure}

\begin{figure}[htbp]
\begin{tabular}{ll}
 a) \includegraphics[angle=0,width=0.2\textwidth]{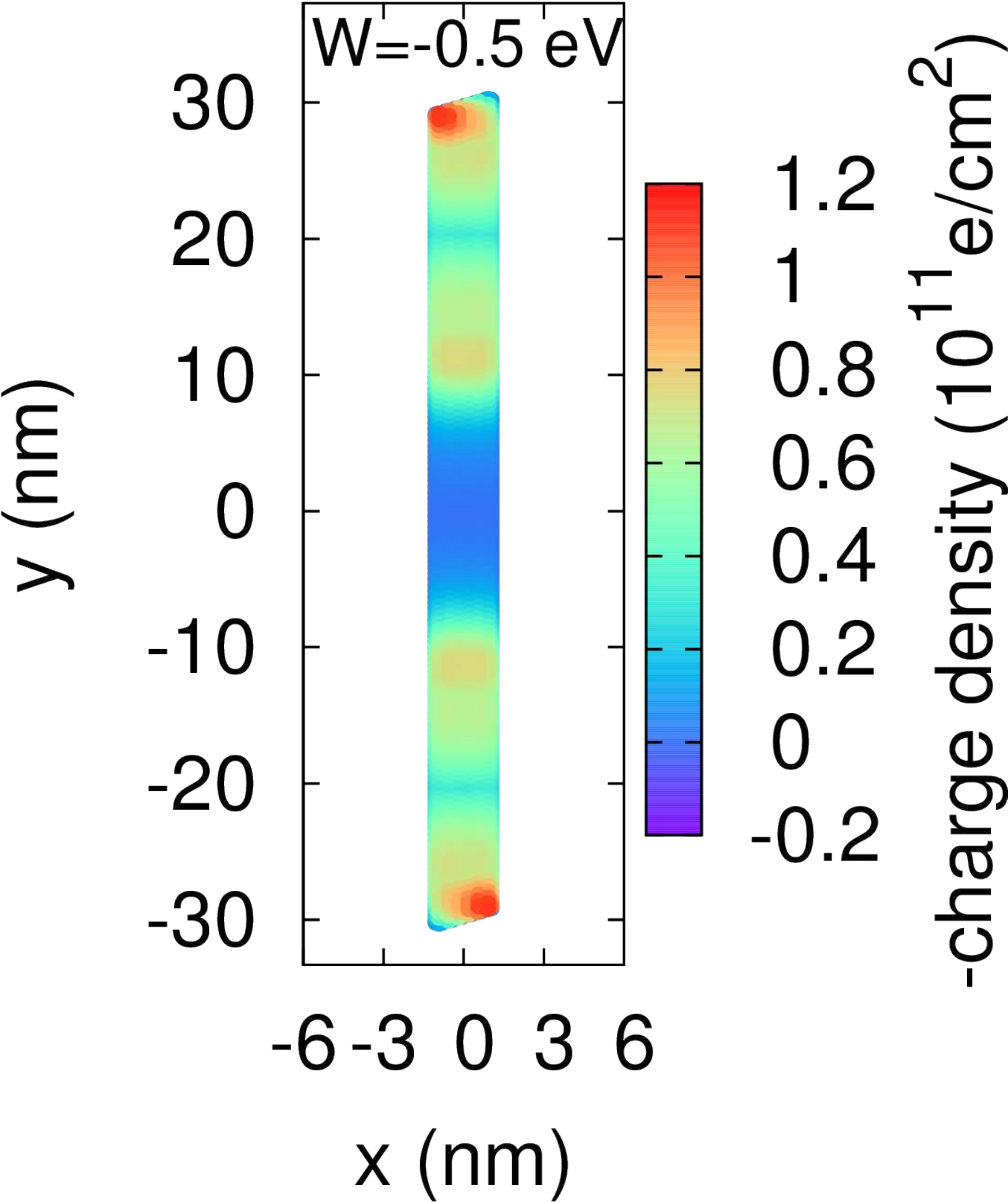} & b) \includegraphics[angle=0,width=0.2\textwidth]{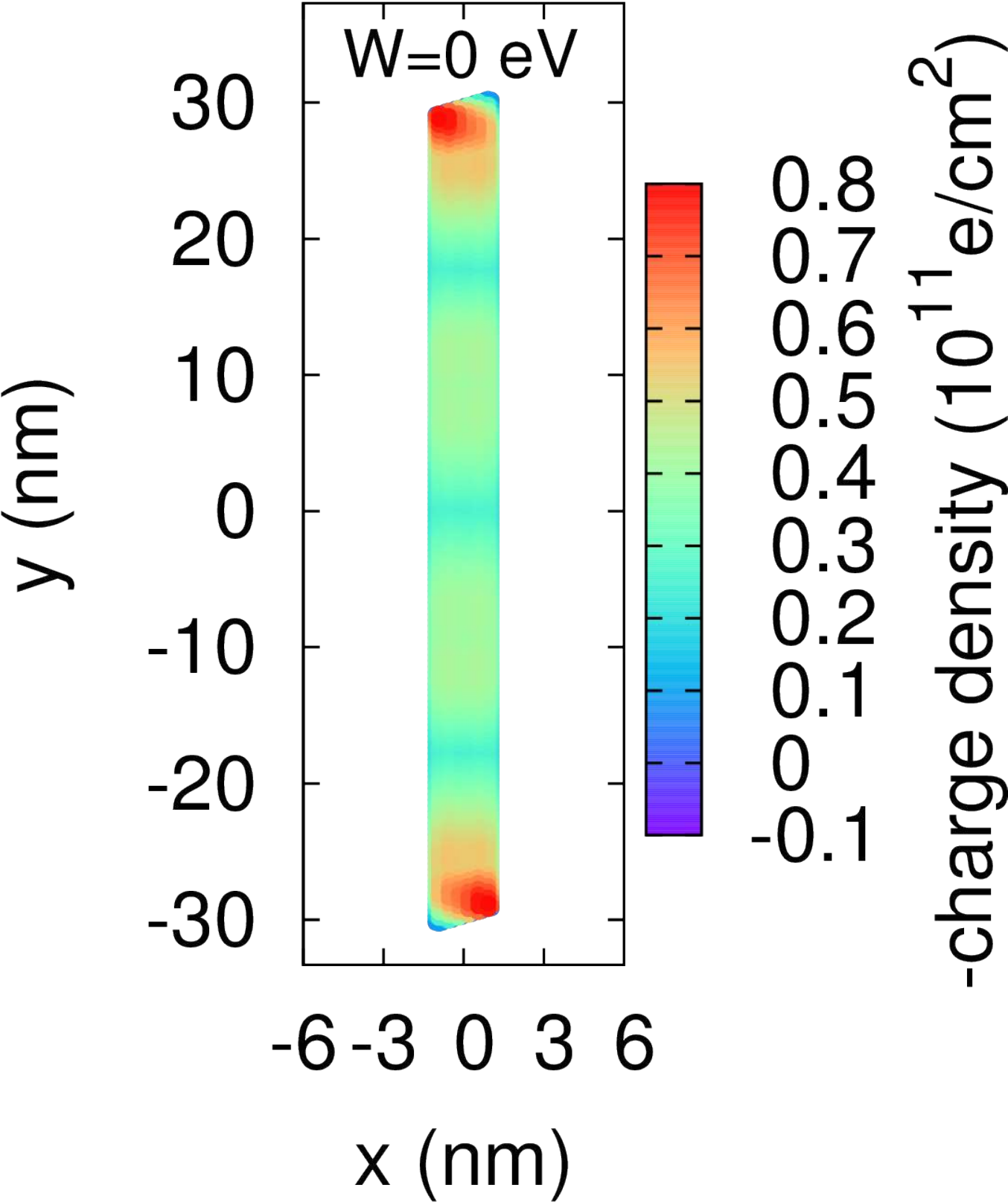}  \\
 c) \includegraphics[angle=0,width=0.2\textwidth]{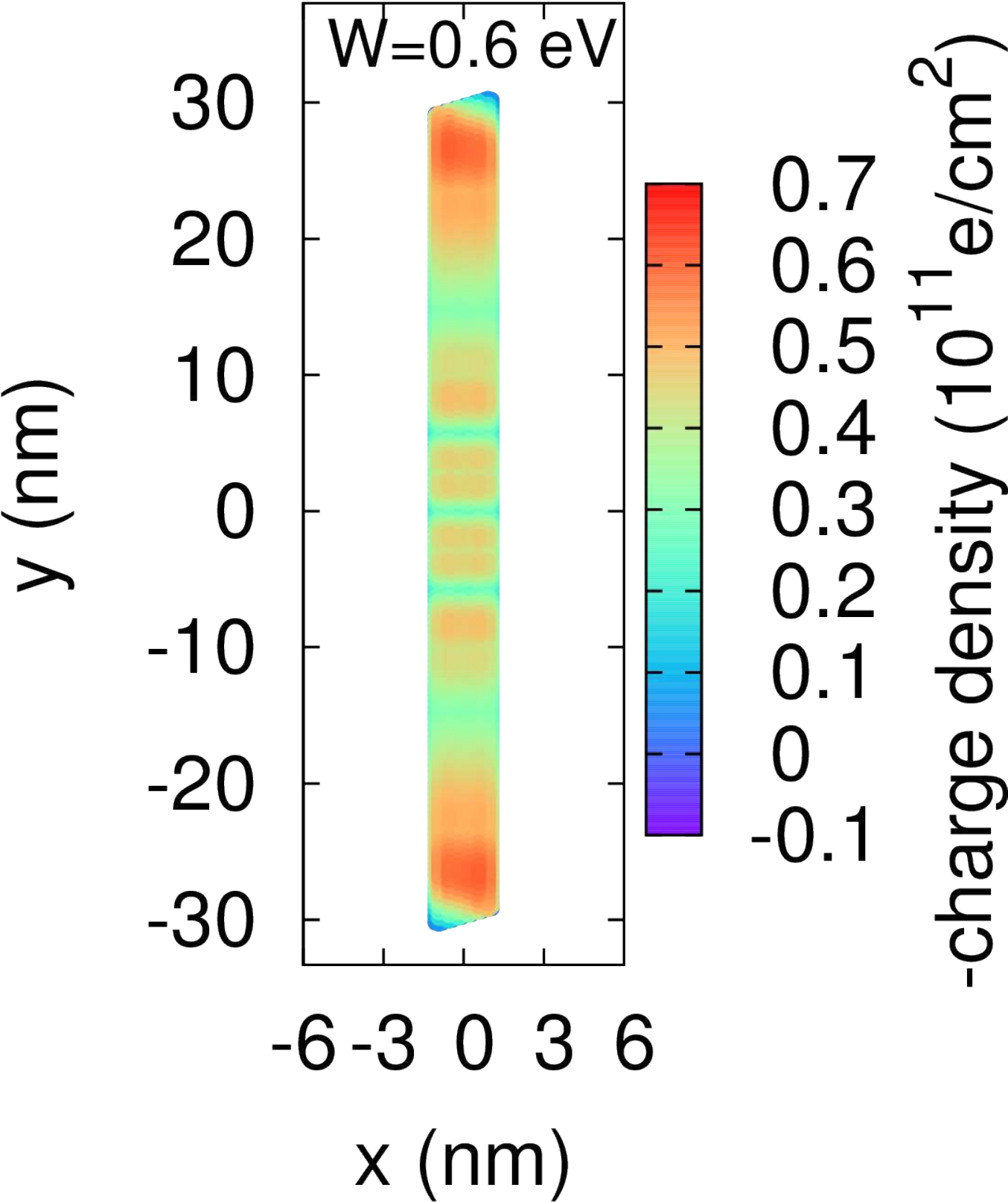} & d) \includegraphics[angle=0,width=0.2\textwidth]{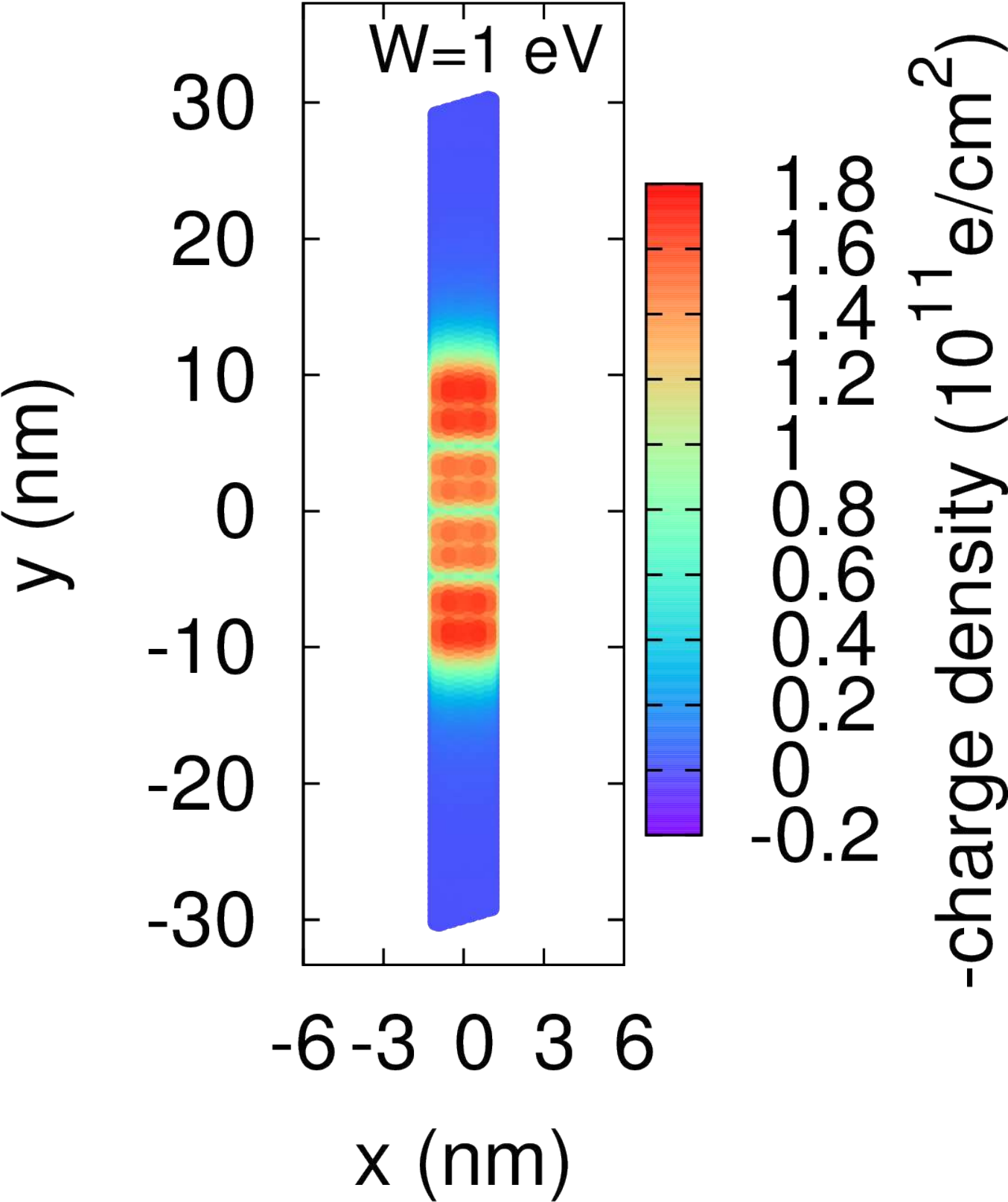} \\
 e) \includegraphics[angle=0,width=0.2\textwidth]{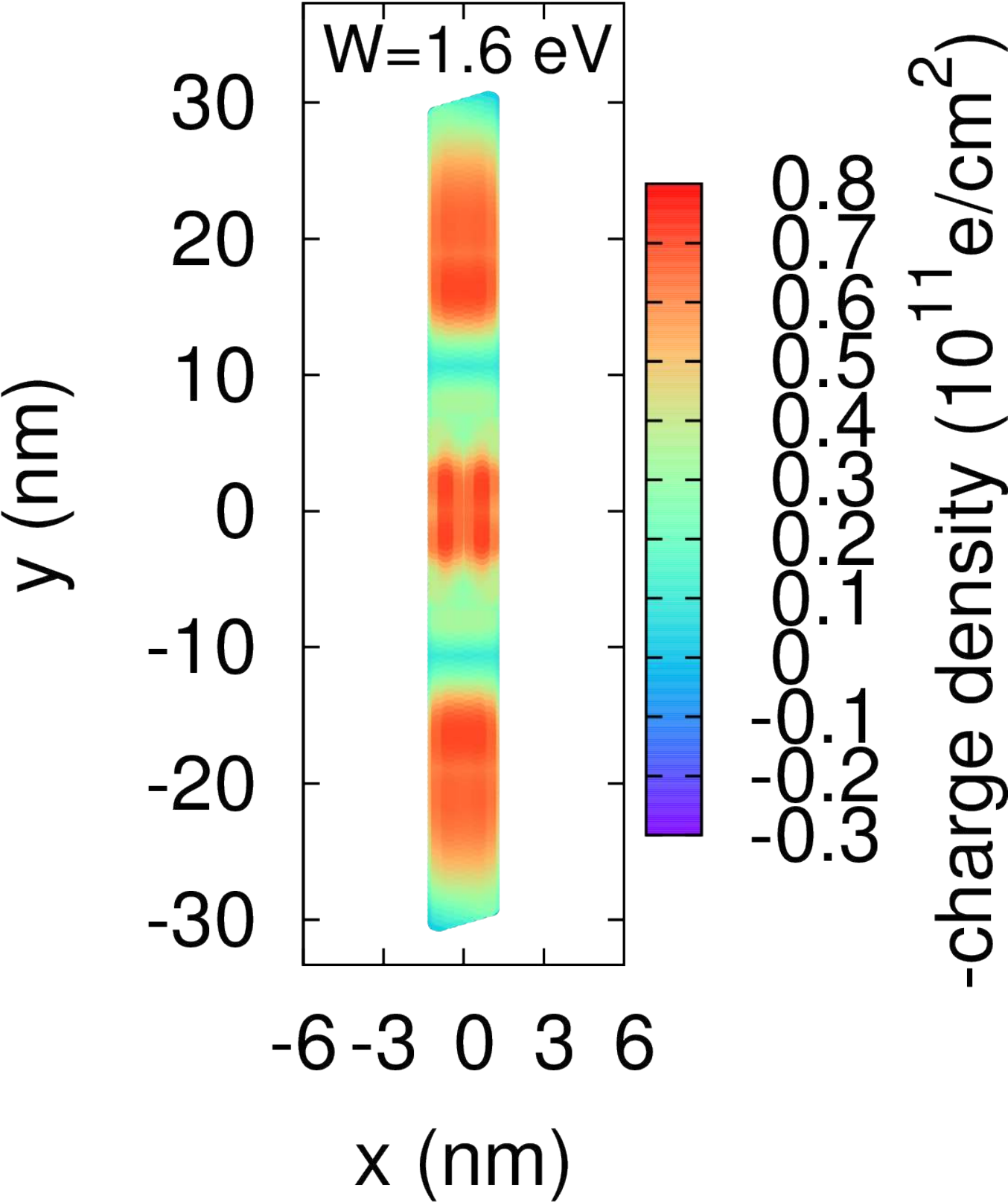} & f) \includegraphics[angle=0,width=0.2\textwidth]{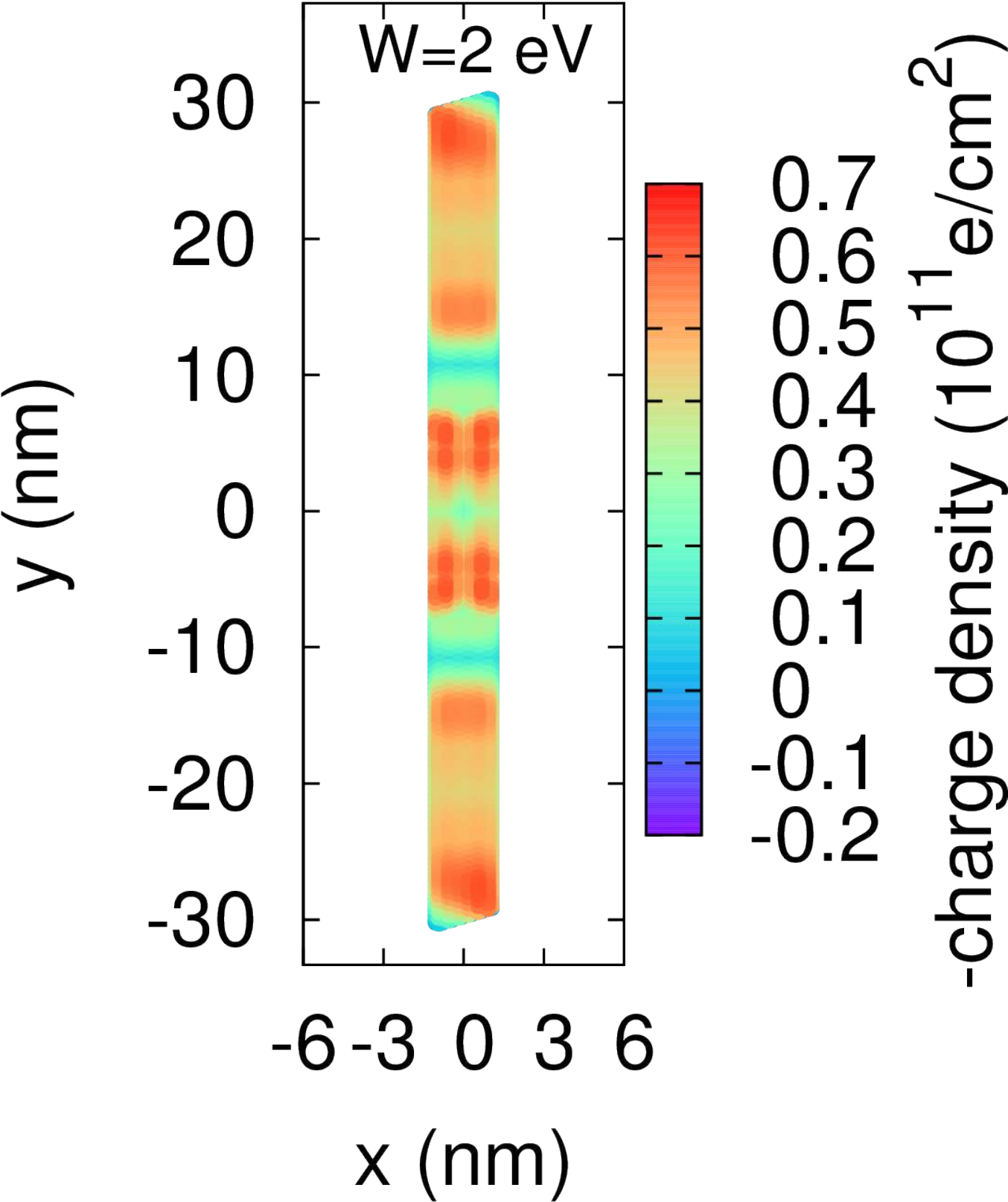}\\
\end{tabular}
\caption{The charge density of 8-th excess electron added to the system calculated as a difference of the total charge for $N_e=8$ and $N_e=7$ for the ribbon of Fig. 3(a).
The $\mu_8$ line is marked by the thick solid line in Fig. 3(a).}
%{\color{red}: prosze wpisac $W$ na rysunkach. }}
 \label{osmy}
\end{figure}

\subsection{Charging the nanoribbon quantum dot}

In Fig. \ref{flejkcz}(a,c) we plotted the chemical potentials for the homogenous nanoribbon with $18$ [Fig. \ref{flejkcz}(a)] and $36$ [Fig. \ref{flejkcz}(c)] atoms across the channel.
The considered number of excess electrons varies between $N_e=-24$ and $N_e=+24$.
For a system in contact with an external electron reservoir the ribbon will be charged by electrons up to the electrochemical potential of the
reservoir $\mu_r$, i.e. the ribbon will contain $N$ electrons for $\mu_{N+1}\ge \mu_r \ge\mu_N$, \cite{tarucha}. A necessary condition for the current flow at low bias is
$\mu_r=\mu_N$, otherwise the charge transfer across the system is in the Coulomb blockade regime.
The horizontal green line in the Fig. \ref{flejkcz}(a,c) marks
$\mu_r=-803.92$ meV, which is the value of the Fermi energy calculated for the neutral ribbon $N_e=0$. This value -- as we find -- is independent of the size of the ribbon
and corresponds to the center of the gap at the chemical potential spectrum for $W=0$, and can be treated as the charge neutrality point for the interacting system.
The chemical energy gap -- corresponding to the gap in the transport experiments \cite{han,chiu,stampfer,xliu,nrb3} -- for the ribbon with 36 atoms across is increased
from 268.5 meV [Fig. \ref{ribo}] to 304.7 meV [Fig. \ref{flejkcz}(c)] by the electron-electron interaction.
The chemical potential spectra -- calculated with the restricted basis -- are perfectly symmetric with respect to the point inversion across the center of the energy gap.
 The electron-hole symmetry found in this result is quite remarkable, given that the frozen orbitals correspond entirely to the valence band.

Note, that the present calculation deals only with the ground state of the electron system with $N_e$ excess
electrons within the dot.  The gap in Fig. \ref{flejkcz}(a) appears in the chemical potentials
for varied electron numbers and not between the excited states of the single-electron or quasiparticle spectrum \cite{gap1,gap2,gap3}.
The chemical potential gap determines the charging properties of the system and not -- for instance --
the optical ones \cite{Yamamoto,Guclu} for a fixed confined charge. %The quantum dots considered in this paper are small, and we do not find any signatures of the
%spin polarization of the system. The ground-state has a non-zero spin only for odd electron numbers.

In the experimental conditions the center of the chemical potential gap can be set above or below the electrochemical potential of the external reservoir $\mu_r$ depending on the back gate potential.
For $\mu_r$ below (above) this value, the ribbon outside the QD displays a $p$-type ($n$-type) conductivity.
On the other hand the external potential \reff{Vext} forms a confinement for conduction band electrons for $W>0$
or for valence band electrons ($W<0$). For $\mu_r>-803.92$ meV and $W<0$ we have n-type conductivity in the ribbon
outside the quantum dot and confinement of the electrons of the valence band ($p$-type) within the quantum dot. This case is labeled
by 'NPN' in Fig. \ref{flejkcz}(a,c,e,g). Other cases of confinement are denoted in a similar manner.

%Note, that for $W=0$ the chemical potential gap 0.677 eV {\color{red} to jest wartosc dla 18, nie dla 36,
%ile jest dla 36} is much larger than in the energy gap for the non-interacting tight binding spectrum (263 meV).
%{\color{blue} jak wzrost przerwy zalezy od szerokosci ukladu, czy w ogole?,
%komentarz literaturowy potrzebny}.
%The electron-electron interaction increases the energy gap in the chemical potential spectrum at $W=0$, since with the first excess electron
%in the nanoribbon a double occupancy of the $p_z$ carbon orbitals appears.

The electrons added to the nanoribbon fill the energy levels
up to the electrochemical potential of the external reservoir $\mu_r$,
but do not necessarily charge the quantum dot itself.
The charge confined in the quantum dot $Q_d$ -- for the nanoribbon with 18 and 36 atoms across the channel-- as a function of $W$ is plotted in Fig. \ref{flejkcz}(b,d) for $N_e=-24,-23,\dots 24$.
The value of $Q_d$ is found by integration of the
electron charge within the $[-1.2R,1.2R]$ segment of the nanoribbon upon subtraction
of the positive ions present within this region.
%In Fig. \ref{dftrib}(a,b) we marked two lines 0 and -803.92 meV which correspond to -- set arbitrarily to fix attention -- the electrochemical potential $\mu_r$ of an electron reservoir external to the nanoribbon.
With the green (red) line
we plotted the number of excess electrons confined inside the dot for $\mu_r=0$ and $\mu_r=-803.92$ meV.
The number of confined electrons jumps in steps between the lines corresponding to a given
number of electrons {\it within the nanoribbon}.
For $\mu_r=0$ -- the value above the chemical potential gap -- the entire nanoribbon
is filled with electrons, over the potential of the dot -- see the red line Fig. \ref{flejkcz}(b,d)
-- and the charge within the dot takes nearly continuous and not quantized values.
 On the other hand for $\mu_r=-803.92$ meV -- the center of the gap for $W=0$, the number of electrons confined within the dot takes
 only integer values (green line in Fig. \ref{flejkcz}(b,d)) -- and corresponds to charging the quantum dot with subsequent electrons.
 Note, that the entire spectrum of chemical potentials in Fig. \ref{flejkcz}(a,c) is slanted,
 i.e. the horizontal green line in Fig. \ref{flejkcz}(a,c) gets closer to the valence
 (conduction) band for $W>0$ ($W<0$).
 As the number of electrons inside the nanoribbon grows,
 so does the potential of the background electron charge, hence
the slope of the spectrum, shifting up the center of the gap  for $W>0$.

For $W=0.5$ eV the chemical potentials of $N_e\in(2,5)$ excess electrons appear within the gap of the 36-atoms wide ribbon of Fig. \ref{flejkcz}(c).
The chemical potentials within the gap acquire a stronger dependence on $W$ and the charge within
the dot for these electron numbers [inset in Fig. \ref{flejkcz}(d)] becomes locally independent of $W$.
For $W=0.8$ eV the chemical potentials of $N_e\in[6,9]$ electrons appear inside the gap [Fig. \ref{flejkcz}(d)], and the charge near
the dot approaches a constant value before the end of the ribbon [Fig. \ref{qdrd}(b)]. For $N_e<6$
the quantum dot attracts the electrons from the entire nanoribbon and for $N_e>9$ the excess electron charge is spread all along the system.
In order to inspect the correspondence between the charge localized near the dot, the excess charge in the nanoribbon
and the position of the chemical potential within the gap in Fig. \ref{flejkcz}(c) we displayed
 in Fig. \ref{qdrd}(a,b) the charge within the $[-R_d,R_d]$ segment of the nanoribbon for a varied number of excess
electrons as a function of $R_d$ for potentials $W=0.5$ eV and $W=0.8$ eV, respectively.
For low $N_e$ we find that the charge in the neighborhood of the dot $Q_d$ exceeds the number of excess electrons
(see Fig. \ref{qdrd}(a) for $N_e=1,\dots 4$ and Fig. \ref{qdrd}(b) for $N_e=1,\dots 7$),
and a local maximum of $Q_d$ as a function of $R_d$ is formed.
The excess electron density for $N_e=1$ and $W=0.5$ eV is displayed in Fig. \ref{pcjaly}(a),
with the total potential plotted in Fig. \ref{pcjaly}(b).
The external potential defining the quantum dot attracts also the electrons occupying the states of the valence band.
The electron charge within the dot is accumulated at the expense of the neighborhood of the nanoribbon which acquires a positive charge
-- see Fig. \ref{pcjaly}(c).
For $N_e=1$ the positive background is left along the entire ribbon up to its
slanted ends (Fig. \ref{qdrd}(a)).
%Nevertheless, the potential within the nanoribbon tends to a constant value outside the quantum
%dot [Fig.\ref{pcjaly}(b,c)] indicating the screening of the external potential by the electron system.
For $W=0.5$ eV the local maximum of $Q_d(R_d)$ is no longer present for $N_e=5$ [Fig. \ref{qdrd}(b)].
For $N_e=5$ the excess charge is present only within the quantum dot [Fig. \ref{pcjaly}(d)].
Note, that for $N_e=5$ the electron system spontaneously forms barriers in the total DFT potential 
separating the quantum dot from the rest of the system [Fig. \ref{pcjaly}(f)].
The appearance of the barriers explains the saturated excess charge observed as a function of $R_d$ in Fig. \ref{qdrd}(a).
For $N_e>5$ the value of $Q_d$ grows monotonically with $R_d$
up to the ends of the nanoribbon [see Fig. \ref{qdrd}(b)],
indicating that the excess electrons do not fit inside the quantum dot [see Fig. \ref{pcjaly}(g) for $N_e=9$] and
are only confined due to a finite length of the nanoribbon.
In this case the total potential is no longer constant outside the quantum dot [Fig. \ref{pcjaly}(h,i) for $N_e=9$].

Figure \ref{wt}(a) shows the charge localized within the quantum dot for a fixed $W$ as a function
of $\mu_r$. The values plotted by thick lines correspond to the chemical potential in the gap of Fig. \ref{flejkcz}(a).
Only within the gap the charge acquires quantized integer values.

In the experiment \cite{xliu} the transport gap is observed for both the back gate potential and the top gate potential
-- see the black cross in Fig. 1(c) of Ref. \onlinecite{xliu}. The gap as a function of the back gate potential \cite{xliu} corresponds to the gap observed in Fig. 3(a,c,e,f) as a function of the chemical potential
(or $\mu_r$).
We marked this gap by the solid orange lines in Fig. 3(a).
The transport gap related to the top gate potential [the diagonal black region in Fig. 1(c) or Ref. \onlinecite{xliu}] is also present in our results.
The gap due to the top gate corresponds to an extension of the central gap region in the direction parallel to the slopes of chemical potentials
for the states localized inside the quantum dot [dotted orange lines in Fig. \ref{flejkcz}(a)].
Let us focus on the chemical potential of 8 excess electrons [bold black line in Fig. 3(a)] as a function of the gate voltage and look at the
location of the 8th electron added to the system as a function of the external (top gate) voltage.
For this purpose we subtracted the total charge densities for $N_e=8$ and $N_e=7$ systems for a varied number of electrons
with the results plotted in Fig. \ref{osmy}.
For $W=-0.5$ eV [Fig. \ref{osmy}(a)] we are in the center of the gap marked by the dashed orange lines in Fig. 3(a).
The 8th electron when added to the system occupies the ribbon connections and not the quantum dot. The flow of the current
across the dot is then blocked, hence the vanishing current found in Ref. \onlinecite{xliu} for the transport gap defined by the top-gate potential.
Note, that $\mu_8$ within this gap have a negligible dependence on $W$ - which also indicates that the 8th electron is localized outside the quantum dot.
When we leave the transport gap [Fig. \ref{osmy}(b,c) for $W=0$ and $W=0.6$ eV] the 8-th electron occupies both the ribbon connections and the quantum dot,
allowing for the flow of the current across the system. In this region the chemical potential acquires a pronounced dependence on $W$. For $W=1$ eV [Fig. \ref{osmy}(d)] we are in the center of the gap on the Fermi energy scale (back gate of Ref. \onlinecite{xliu}) and the 8-th electron occupies the dot only and the variation of $\mu_8$ with the external potential is the strongest.
In the region marked by the orange solid lines in Fig. 3(a) the transport across the quantum dot can occur via the dot localized states \cite{chiu,stampfer,moreview,droescher}, naturally provided that the ribbon connections have non-zero density of states in this energy range.
When we leave the transport gap [Fig. \ref{osmy}(e,f) for $W=1.6$ eV and 2 eV] the electron occupies both the dot and the ribbons allowing for the electron flow across the system.

The transport gap due to the top gate potential in Fig. 1(c) of Ref. \cite{xliu} depends strongly on the back gate potential.
An increase of the top gate potential in Fig. 1(c) of Ref. \cite{xliu} corresponds to an increase of $W$.
A more negative back gate potential in Fig. 1(c) of Ref. \cite{xliu} corresponds to a more negative values of the chemical potentials in
Fig. \ref{flejkcz}(a). Hence for large positive $W$ the gap appears for large negative chemical potentials in Fig. \ref{flejkcz}(a) in a correct qualitative correspondence with Fig. 1(c) of Ref. \cite{xliu}, where
the transport gap for large positive top gate potential appears for large negative back gate potentials.

\subsection{Charging the QD induced within the graphene flake}

Figure \ref{flejkcz} compares the chemical potential spectra and the charge
localized within the quantum dot for a nanoribbon containing 18 atoms across the channel
with [Fig. \ref{flejkcz}(e,f)] and without [Fig. \ref{flejkcz}(a,b)] a hexagonal flake
defined in the center of the ribbon. For the quantum dot induced within the flake the chemical potential gap as a function of $W$ is much thinner
as in the results obtained for non-interacting electrons cf. Fig. \ref{ribo}(a) and Fig. \ref{ribo}(c).
The dependence of the charge confined in the quantum dot as a function of $W$ is displayed in Fig. \ref{flejkcz}(f).
For a fixed value of $W$ the flake contains a larger number of electrons than for the ribbon QD [cf. Fig. \ref{flejkcz}(f) and Fig. \ref{flejkcz}(b)].
Nevertheless, no excess charge appears in the flake for a neutral system $N_e=0$ at $W=0$ [Fig. \ref{flejkcz}(f)].
In contrast to the nanoribbon QD -- for $N_e>0$ a few extra electrons are localized within the flake QD also at $W=0$.
In consequence, for the flake the central band of a quantized confined charge in Fig. \ref{flejkcz}(f) is more or less constant as a function of $W$ and $N_e$
in contrast to the nanoribbon [Fig. \ref{flejkcz}(d)] where for $W=0$, $Q_d$ is only integer for $N_e=0$.
The plots of the confined excess charge as a function of the cutoff-radius $R_d$ for the flake quantum dot given in Fig. \ref{qdrd}
indicate a large flexibility of the confinement potential in terms of the number of confined electrons. For $W=0.8$ eV  [Fig. \ref{flejkcz}(f)]
a saturation of the excess charge is obtained for $N_e\in(3,14)$ near $R_d\simeq 10$ nm.
%The cross section ($y=0$) of the excess charge density and the DFT potential are plotted in Fig. \ref{pcjalyh} for $W=0.5$ eV and $N_e=4$ excess electrons.
%The DFT potential forms barriers at the exit from the channel.

In order to estimate the effect of the finite size of the ribbons in the system we plotted in Fig. \ref{flejkcz}(g) and \ref{flejkcz}(h)
the results for the hexagonal flake embedded in a longer ribbon. The total length of the system was increased from $L=60.1$  nm - as in Fig. \ref{flejkcz}(e,f) and elsewhere in this work,
to 90 nm. We observe that the chemical potentials within the energy gap and the width of the energy gap are unchanged. Nevertheless, the chemical
potentials  which are outside of the gap are shifted towards the neutrality point -- the band of chemical potentials for $N_e\in [-24,+24]$ electrons becomes thinner
on the vertical scale. The longer ribbons reduce the interaction energy for the carriers within the ribbons, and the chemical potential approach the continuum limit.
Concerning the charge confined within the dot displayed in Fig. \ref{flejkcz}(h) -- as compared to Fig. \ref{flejkcz}(f) we can see that the central part of the figure
-- corresponding to a quantized charge within the dot -- is left unchanged. Outside the energy gap the entire nanoribbon contains a larger number of electrons for a
given $\mu_r$ and the line for $\mu_r=0$ -- as compared to Fig. \ref{flejkcz}(f) -- tends to acquire a continuous dependence on $W$.
Same conclusions are reached for the charge localized within the flake quantum dot for a fixed $W$ as a function of $\mu_r$ [cf. the solid and dashed
lines in Fig. \ref{wt}(b) corresponding to $L=90$ nm and $L=60$ nm, respectively].
%Note, that both the transport gaps as obtained with both values of $L$ [Fig. \ref{flejkcz}(e,g)] have a very similar position and width.

\subsection{Frozen valence band charge}
\begin{figure}[htbp]
\begin{center}
\begin{tabular}{ll}
a) &\includegraphics[angle=0,width=0.4\textwidth]{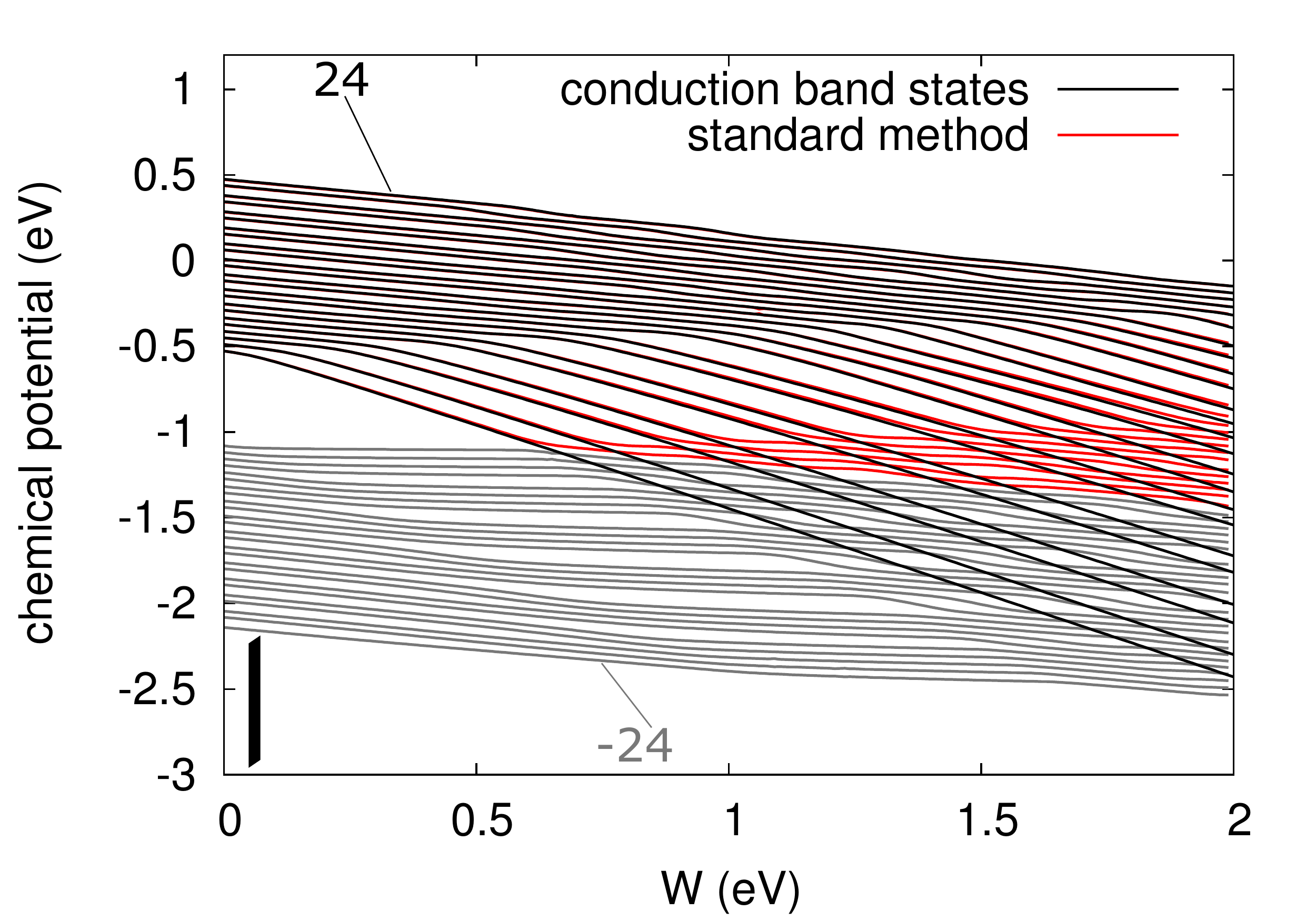}\\
b) & \includegraphics[angle=0,width=0.4\textwidth]{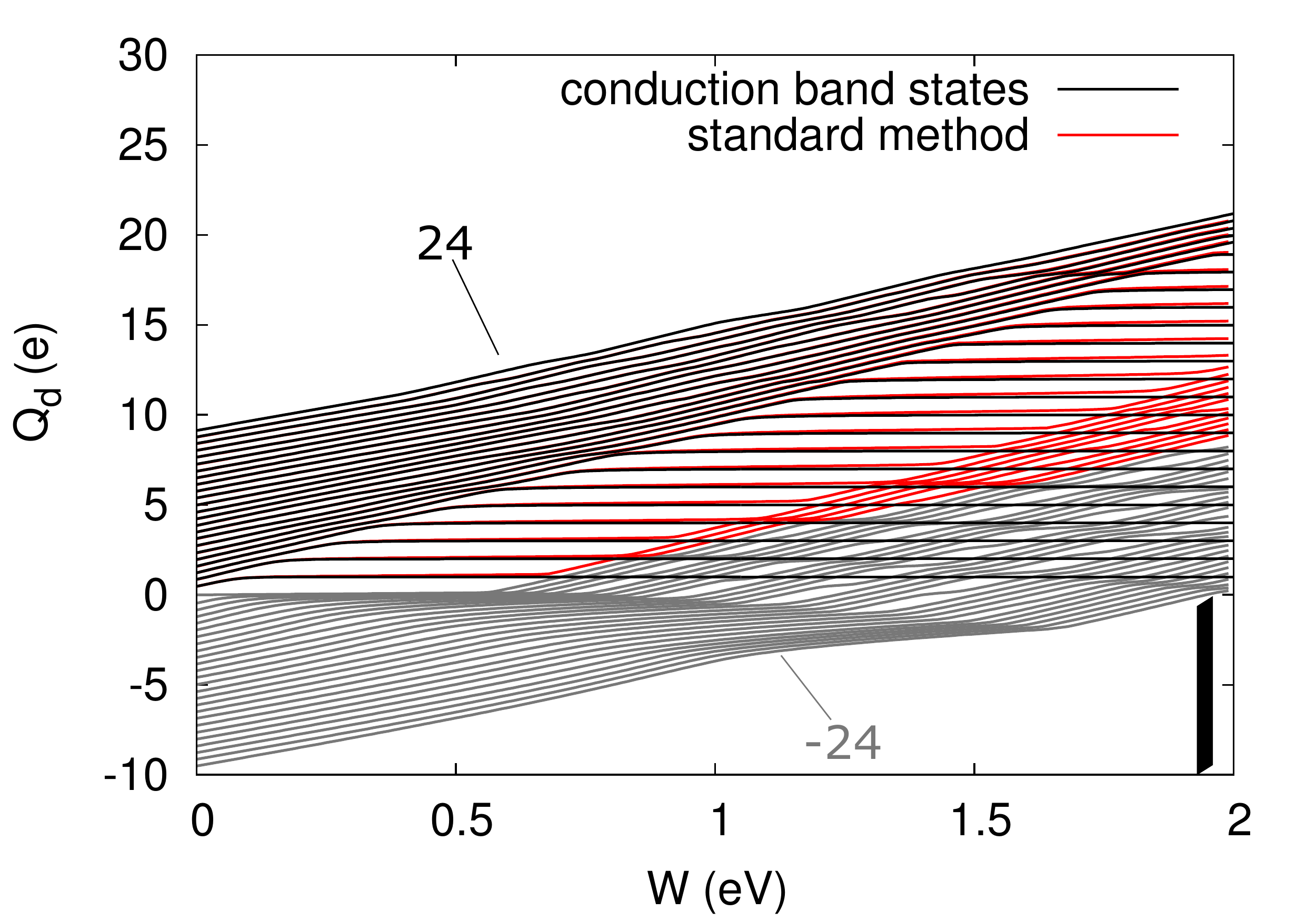}
\end{tabular}
\caption{(a) Chemical potentials for $N_e=-24,\dots 24$ excess electrons
for the ribbon with 18 atoms across the channel as calculated by the standard
method of the present work (red lines for $N_e=1,\dots 24$ and grey lines for $N_e=-24,\dots -1$) and with the basis of conduction band only (black lines),
with all the single-electron states below the energy gap considered as frozen.
(b) Charge confined within the quantum dot for both approaches as a function of $W$.}
\label{neutrality}
\end{center}
\end{figure}

%\begin{figure}
%\begin{tabular}{ll}
%\includegraphics[angle=0,width=0.4\textwidth]{porall.pdf}
%\end{tabular}
%\caption{
%The charge localized in the quantum dot $Q_d$ of a 18-atoms wide nanoribbon calculated
%with the exact approach (thin lines) and with the basis limited to the conduction band
%states (thick lines) with frozen occupied valence band states. } \label{wtn}
%\end{figure}

The results presented so far were obtained within the method in which both the conduction and valence band states
formed the basis for solution of LSDA equations. Let us suppose that we are only interested in the conduction band
states, or charging the quantum dot with electrons $N_e>0$ and not holes -- which appears for $W>0$.
Let us discuss the applicability of the method with the basis limited to conduction band states -- for which
the complexity of the calculations can be significantly reduced.
For this purpose we {\it i)} diagonalize Hamiltonian (1), {\it ii)} fix the charge density (4) as obtained
for a charge neutral system taking the summation up to $j=-1$ in Eq. (4), then {\it iii)} the LSDA equations
are solved in the basis given by Eq. (3) but with summation starting from $j=1$.
Since the Hamiltonian (1) does not contain any potential, the frozen charge density (4) of all the occupied valence band states below the energy gap is uniform and the same
on each atom of the crystal lattice. Then, the solution of the DFT equations is performed in the conduction band only,
with the frozen valence charge producing only a constant potential shift, due to the exchange-correlation component of the Perdew-Zunger
potential.

The results for the chemical potentials for a nanoribbon with 18 atoms across the channel are displayed in Fig. \ref{neutrality}(a).
The basis limited to the conduction band correctly reproduces the chemical potentials above and
inside the gap. No additional shift was necessary to make this data coincide.
As $W$ is increased, the chemical potential spectrum for low $N_e>0$ continues into the
quasi-continuum of the chemical potentials of the valence band.
The basis limited to the conduction band state predicts that the charge confined within the quantum dot remains
quantized at integer values for large values of  $W$ [black lines in Fig. \ref{neutrality}(b)], while in the exact
calculation [red lines in Fig. \ref{neutrality}(b)] the system enters the energy continuum for the p-type conductivity.
Near the bottom of the gap we find that for a fixed value of $N_e$ the chemical potential calculated with the limited basis goes down on the energy scale as compared to the exact calculation [Fig. \ref{neutrality}(a)].
In this region we find that for a fixed $N_e$ the exact calculation gives values of the quantum-dot-confined charge which are {\it i)} larger [Fig. \ref{neutrality}(b)] than
the ones obtained with the limited basis and {\it ii)} continuous as a function on $W$ in contrast to the quantized values for the limited basis.
This discrepancy is a result of filling the quantum dot by the electrons coming from the valence band from outside the quantum dot, which is present in the exact calculation but neglected
in the limited basis.
In the limited basis the excess charge confined within the quantum dot is underestimated, hence the underestimate of the interaction energy and in consequence lower values of chemical potentials as compared
to the exact calculation [see the red and black lines in Fig. \ref{neutrality}(a) before the black lines enter the gray band of the valence band states].

\subsection{Potential profile and edge disorder}

\begin{figure}[htbp]
\begin{center}
\includegraphics[angle=0,width=0.45\textwidth]{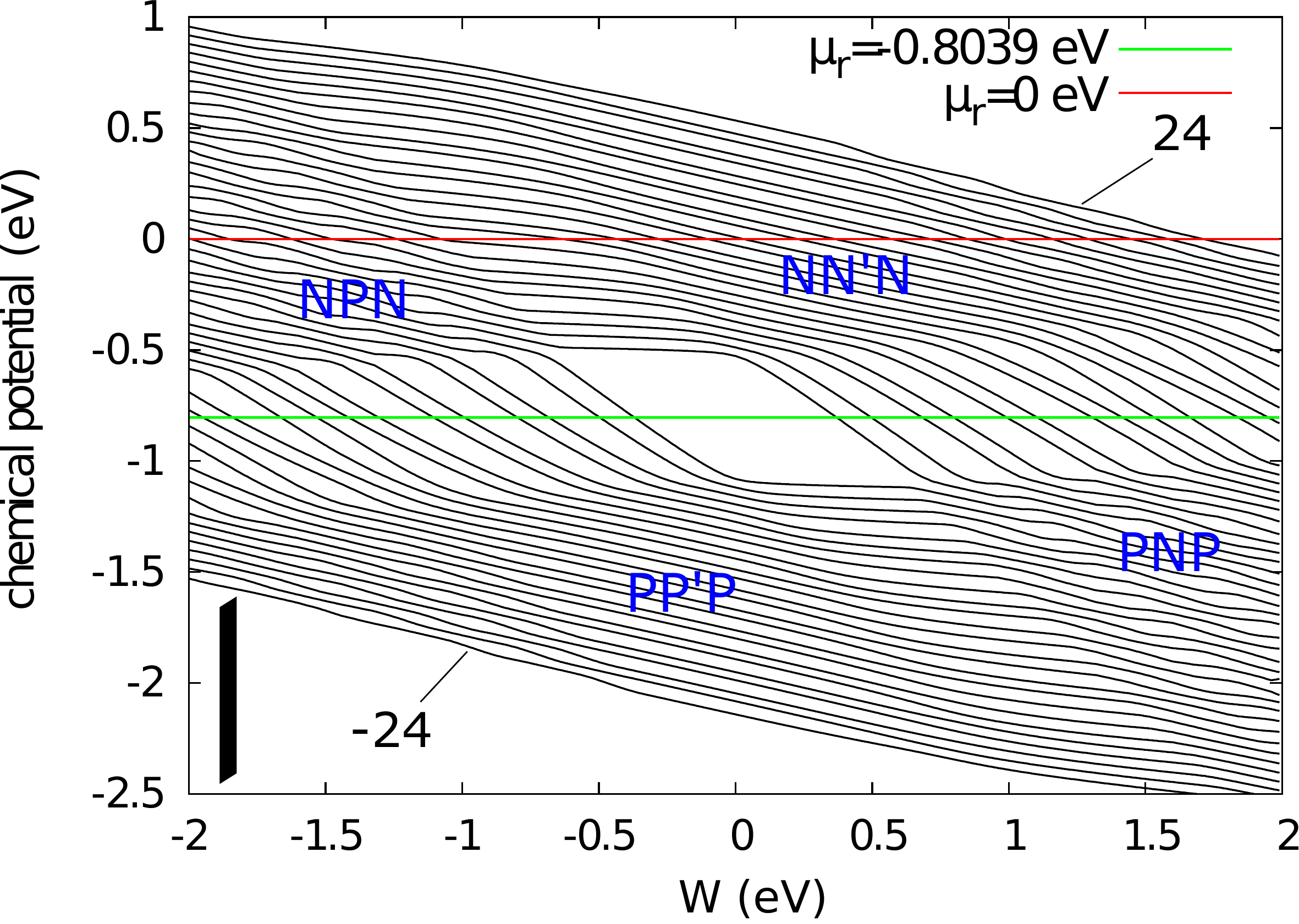}\\
\caption{Chemical potentials for $N_e=-24,\dots 24$ excess electrons
for the ribbon with 18 atoms across the channel for confinement potential $-W\exp\left(-\left(\frac{y}{R} \right)^2\right)$ instead of Eq. (12). For the results with Eq. (12) see Fig. 3(a).}
\label{mod}
\end{center}
\end{figure}

In this subsection we present the results for a modified external potential profile and the disordered edge within the flake.
The choice of the form of the external potential defining the quantum dot [Eq.(12)] was taken arbitrarily, as a smooth short-range potential.
In order to demonstrate that the adopted results remain qualitatively unchanged when the potential profile is varied we performed calculations for a Gaussian confinement $V_{ext}(\mathbf{r})=-W\exp\left(-\left(\frac{y}{R} \right)^2\right)$, 
i.e. for a power of two instead of 4 in the exponent. The results for the chemical potential are displayed in Fig. \ref{mod} to be compared with Fig. 3(a) for the potential given by Eq. (12). The pattern of chemical potentials remains unchanged, and only quantitative differences are observed.

\begin{figure}[htbp]
\begin{center}
\begin{tabular}{lc}
a) & \includegraphics[angle=0,width=0.35\textwidth]{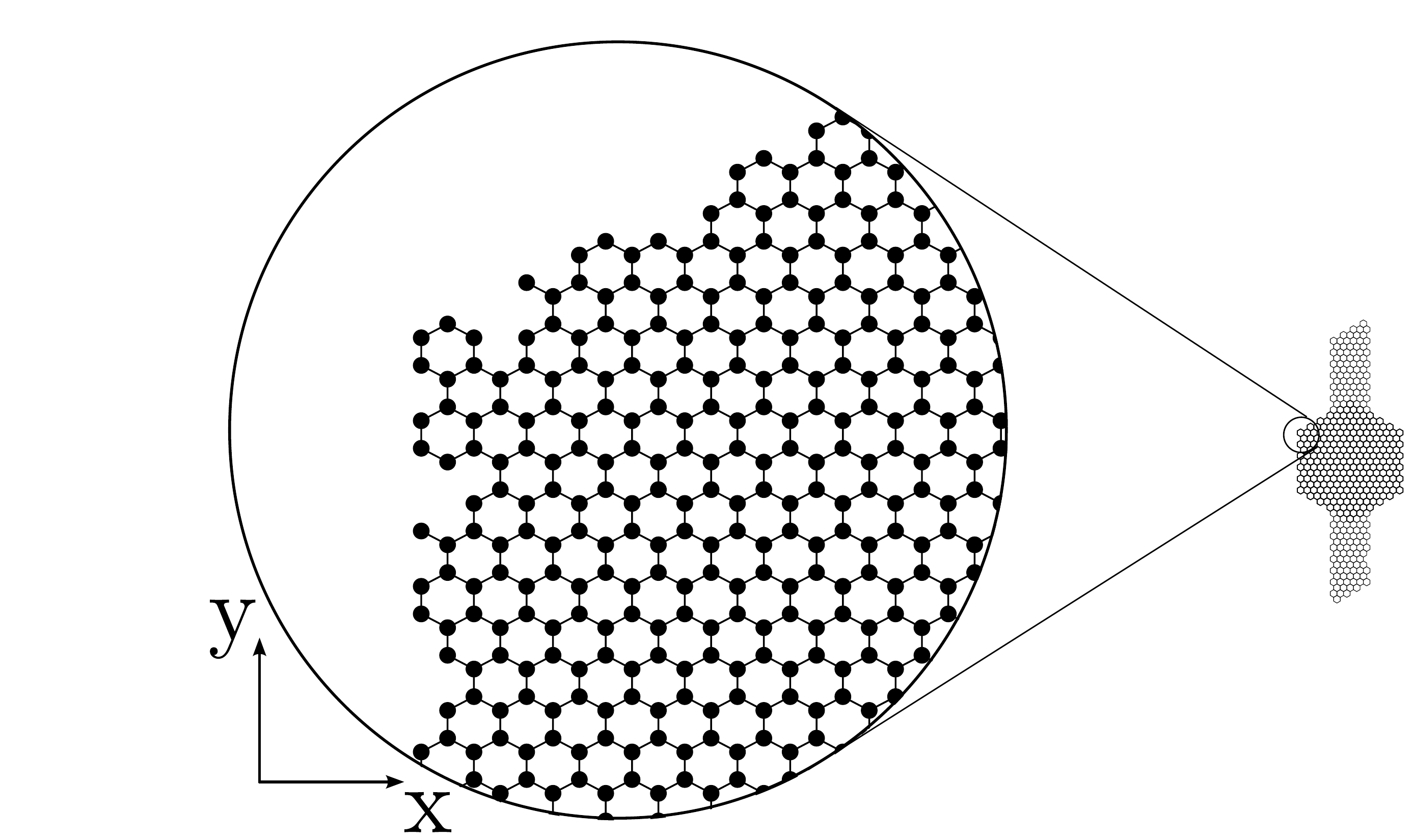} \\
b) & \includegraphics[angle=0,width=0.45\textwidth]{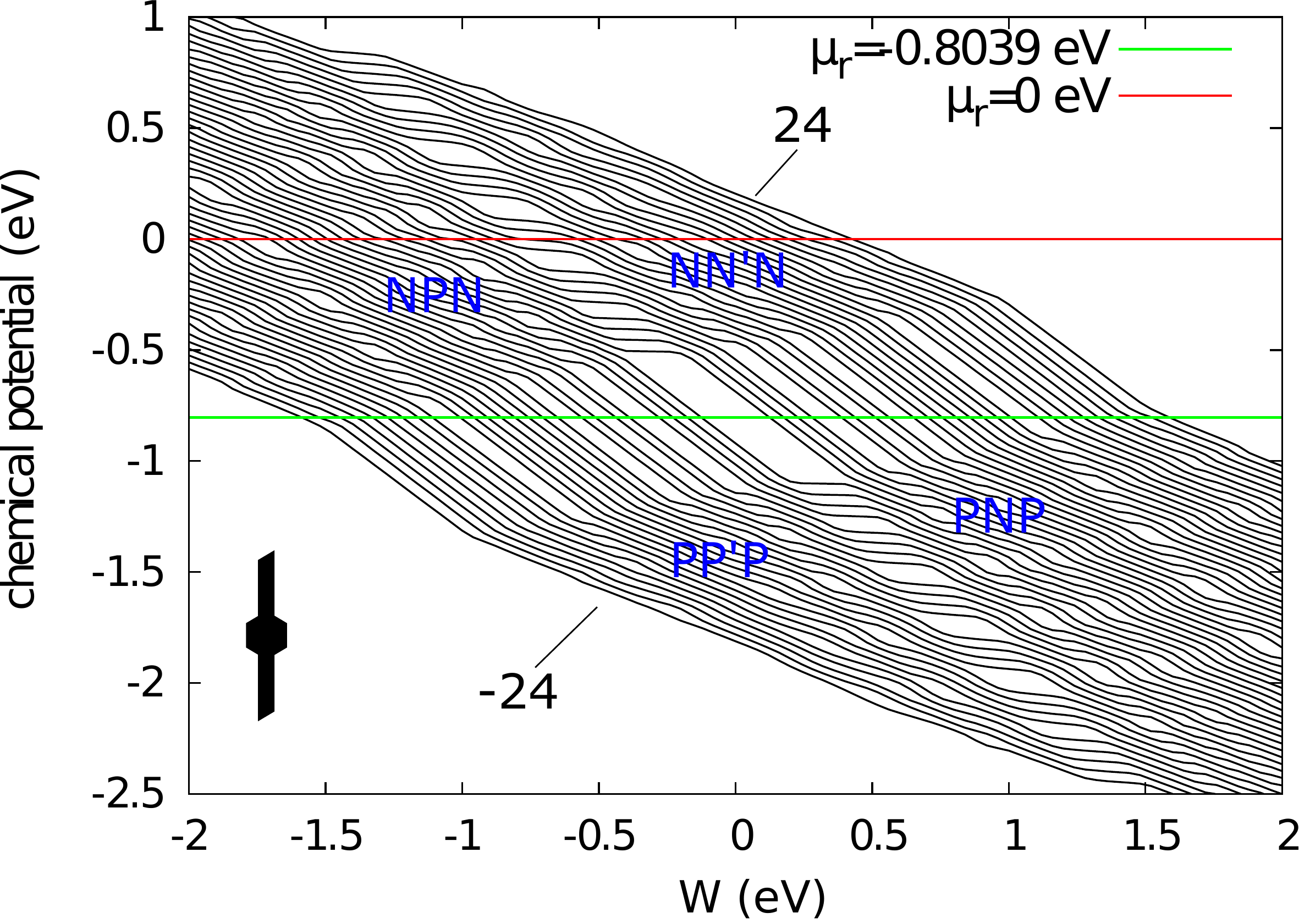} \\
c)&\includegraphics[angle=0,width=0.45\textwidth]{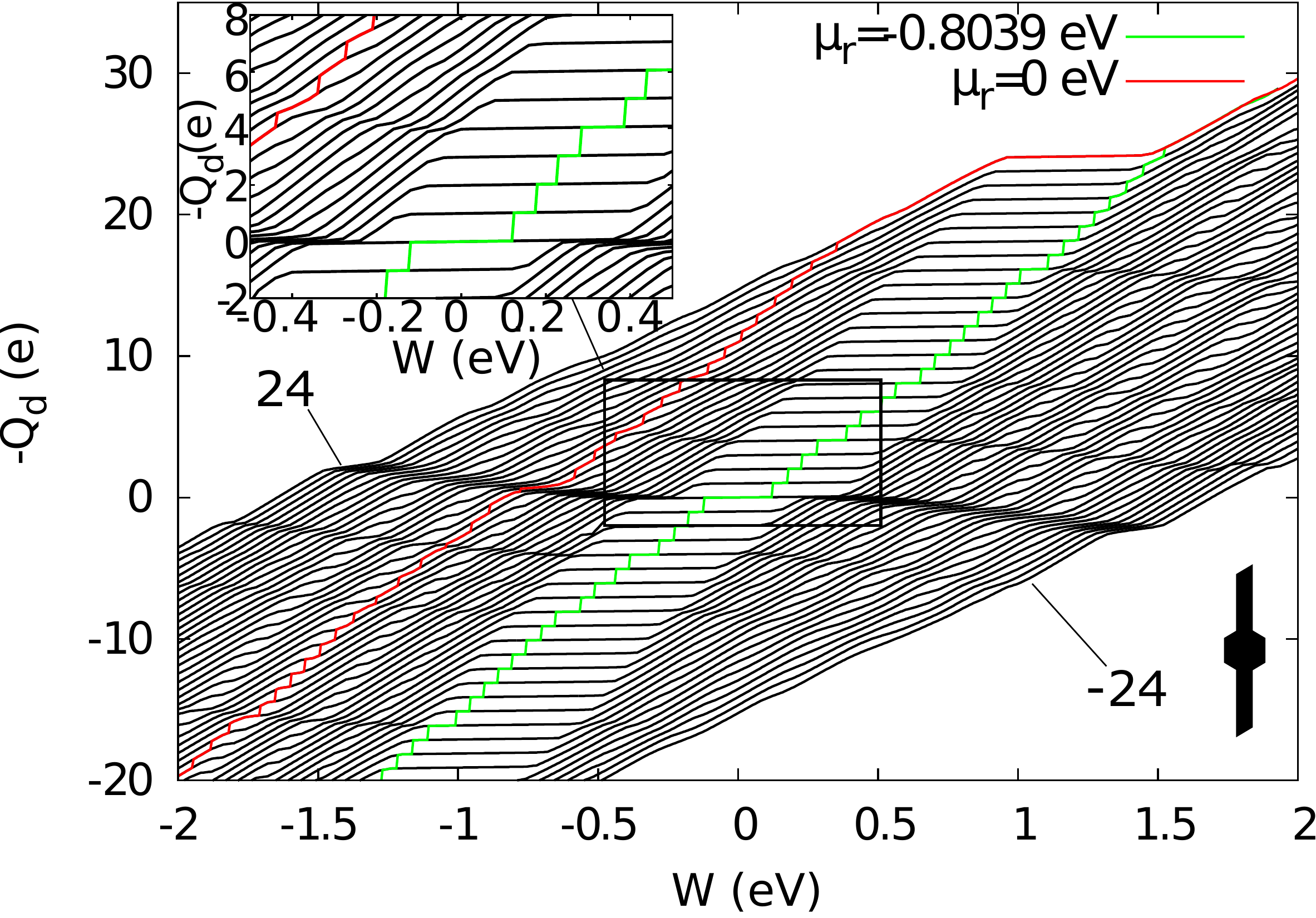} \\
 \end{tabular}
\caption{Same as Fig. 3(c,d) only with disordered edge of the flake containing both armchair and zigzag fragments as displayed in (a), see text. }\label{defectss}
\end{center}
\end{figure}

For the edge disorder effects we took the hexagonal flake with pure armchair edges end removed two consecutive atoms
every 10 of atoms along the edge. The form of the boundary is displayed in Fig. \ref{defectss}(a). The edge contains 
fragments zigzag edge and single separated ions within armchair fragments. The results with ideal armchair edges of the
flake were presented in Fig. 3(b,c) for the chemical potentials and the charge localized within the dot. Again only quantitative differences can be 
spotted with no overall change to the quantitative description of the effect.

\section{Summary and Conclusions}

We have studied the charge distribution in a semiconducting graphene nanoribbon with a quantum dot defined in its center by an external potential.
The calculations were performed within the tight-binding method with the electron-electron interaction introduced by the mean field DFT approach.

We determined the excess charge localized within the dot as a function of the Fermi energy and the external potential within the transport gap and outside, when the entire structure is flooded by carriers.
The calculated chemical potential spectra bear signatures of the transport gaps which agree with the ones observed experimentally as functions of both the back and top gate potentials.
The inline flake embedded in the gated region of the nanoribbon allows for confinement of the excess charge already in the absence of the external potential, however no excess charge within
the flake is found for a charge-neutral system unless an external potential is present.
We found that for the frozen valence band charge the charging of the quantum dot with excess electrons $N_e>0$ can be quite well described for $N_e>0$ above the lower limit of the transport gap.
However, the lower limit of the transport gap is overlooked by the basis limited to the conductance band states.
 Moreover, the limited basis neglects the QD charging with the valence band electrons which appears for large external potentials. The neglect
 leads to an underestimate of both the confined charge and the chemical potential, particularly near the lower limit of the transport gap and below on the Fermi energy / potential scale.

\section*{Acknowledgements}
This work was supported by National Science Centre
according to decision DEC-2013/11/B/ST3/03837 and by
PL-GRID infrastructure.

\end{document}